\documentclass[lettersize,journal]{IEEEtran}
\usepackage{amsmath,amsfonts}
\usepackage{amssymb}
\usepackage{algorithm}
\usepackage[linesnumbered,ruled,vlined,algo2e]{algorithm2e}
\usepackage{array}

\usepackage{textcomp}
\usepackage{stfloats}
\usepackage{url}
\usepackage{verbatim}
\usepackage{graphicx}
\def\BibTeX{{\rm B\kern-.05em{\sc i\kern-.025em b}\kern-.08em
    T\kern-.1667em\lower.7ex\hbox{E}\kern-.125emX}}
\usepackage{balance}
\usepackage{xcolor}
\usepackage{subfigure}
\usepackage{enumitem}
\usepackage{multirow}
\newcommand{\reviewblue}[1]{\textcolor{black}{#1}}
\begin{document}
\title{FedCoT-VQA: A Federated Learning and Unlearning Framework \textcolor{black}{for Chain-of-Thought Planners in VideoQA}}

\author{ Rui Lu,
Tao Ling,
Dan Wang, Qing Li
\IEEEcompsocitemizethanks{
\IEEEcompsocthanksitem Rui Lu is with the Department of Computing, The Hong Kong Polytechnic University, Hung Hom, Hong Kong. (e-mail: ruilu@polyu.edu.hk)
\IEEEcompsocthanksitem Tao Ling was with the Department of Computing, The Hong Kong Polytechnic University.  (e-mail: cstling@comp.polyu.edu.hk)
\IEEEcompsocthanksitem Dan Wang is with the Division of Environment and Sustainability Academy of Interdisciplinary Studies, Hong Kong University of Science and Technology, Clear Water Bay, Hong Kong. (e-mail: wangdan@ust.hk)
\IEEEcompsocthanksitem Qing Li is with the Pengcheng Laboratory, Shenzhen, China. (e-mail: liq@pcl.ac.cn)
}
}
\markboth{FedCoT-VQA: Author Preprint}{FedCoT-VQA: Author Preprint}
\maketitle
\begin{abstract}
Chain-of-Thought (CoT) planners have emerged as an effective design for VideoQA, where a lightweight planner first generates intermediate reasoning steps to guide temporal evidence selection before answer prediction.
This modularity makes CoT-based VideoQA attractive for federated learning, since only the planner side needs collaborative adaptation while the heavy vision-language backbone can remain fixed.
However, in decentralized settings, the planner must not only be trained efficiently across heterogeneous clients but also support later client deletion requests.
This is challenging because deleted-client influence is reflected both in model parameters and the planner's reasoning-trace behavior.

We present \emph{FedCoT-VQA}, a federated learning and unlearning framework for CoT planners in VideoQA.
FedCoT-VQA consists of three modules: planner-side partitioning (PSP), which exposes a compact shared-residual adaptation space for efficient federated training; server-side aggregation (SSA), which aggregates planner-side updates while maintaining a deletion-ready contribution log; and a residual unlearning module (RUM), which approximates the retained-only counterfactual planner through retained-client replay and selective residual correction, without full retraining.
We evaluate FedCoT-VQA in terms of federated training utility, federated unlearning utility, forgetting quality, and efficiency.
Results show that compared to current federated approaches, FedCoT-VQA preserves strong federated training utility, improving grounding quality by up to 4.45\%.
After unlearning, it retains high accuracy and achieves a counterfactual gap of only 7.38\%.

\end{abstract}

\begin{IEEEkeywords}
Video question answering, federated learning, federated unlearning, chain-of-thought planner
\end{IEEEkeywords}

\section{Introduction}

Large language models (LLMs) have made natural language interfaces and multi-step reasoning practical at scale. Vision-language models (VLMs) extend this capability to visual inputs and enable tasks such as \emph{video question answering} (VideoQA), \textcolor{black}{which answers natural language queries about video streams}. 
Compared with static images, video queries often span long time ranges and require aggregating evidence across temporally separated events, making long-horizon reasoning a central challenge. Processing each frame is computationally expensive, but aggressive downsampling would miss critical evidence and produce incorrect outputs.

Recent approaches mitigate this challenge by incorporating \emph{Chain-of-Thought} (CoT) reasoning into VideoQA. Rather than retraining or repeatedly running the entire VLM over long videos, they decompose a query into intermediate steps and use these steps to drive evidence selection, e.g., which clips or frames to inspect, before generating the final response.
This modular design can be trained by updating a lightweight CoT evidence-selection component while keeping the large visual-language backbone fixed, reducing both training and inference cost. However, it also expands the privacy attack surface during training: intermediate reasoning traces and learned evidence-selection policies may encode client-specific viewing patterns, recurring scene content, or memorized details from private videos. Because these artifacts expose what the model attends to and how it reasons, they can be more privacy-revealing than the final answer and are therefore sensitive to share during training.

Federated learning (FL) is a natural fit for training CoT-enabled VideoQA systems as it avoids centralizing raw videos by keeping data on-device and aggregating model updates, e.g., gradients, to train a shared model cooperatively.
However, FL alone does not satisfy deletion and compliance requirements.
In practice, participants may later request removal of their contributions due to contract termination, policy changes, or \textit{right-to-be-forgotten} obligations. Federated unlearning (FU) addresses this need by updating the global model to approximate the counterfactual model that would have been obtained had a specific client never participated.

\reviewblue{In this paper, we design a novel federated learning and unlearning framework,} \emph{FedCoT-VQA}, for the chain-of-thought planner in VideoQA.
To the best of our knowledge, this is the first framework that jointly studies both federated training and federated unlearning for a chain-of-thought planner in video understanding.
Designing such a framework introduces three major challenges.
First, \emph{efficient federated CoT adaptation}: the system should support collaborative training over distributed clients without federating the entire VideoQA pipeline, so the trainable space must remain compact while still preserving the planner's temporal reasoning ability.
Second, \emph{deletion-ready aggregation under client heterogeneity}: client influence is accumulated through round-wise updates and heterogeneous reasoning behaviors, so the server must aggregate planner updates in a way that both supports robust global learning and preserves sufficient contribution structure for later deletion.
Third, \emph{effective unlearning}: after a deletion request, the framework should remove the deleted client's effect not only from model parameters but also from induced reasoning-trace behavior, while retaining utility for the remaining clients and approximating the retained-only counterfactual model without full retraining.

\reviewblue{To address these challenges, FedCoT-VQA follows a modular design.}
In FedCoT-VQA, we first design \textsc{PSP}, which exposes a compact planner-side adaptation space and separates it into shared and residual components, enabling efficient federated training while preserving reasoning quality. We then design \textsc{SSA}, which aggregates the corresponding planner-side updates and maintains a compact server-side contribution log for unlearning. Finally, we design \textsc{RUM}, which reconstructs a retained-only replay trajectory and selectively revises the more deletion-sensitive part of the planner, thereby approximating the retained-only counterfactual model without full retraining.

We evaluate FedCoT-VQA under decentralized VideoQA settings against both federated training and federated unlearning baselines. The results show that FedCoT-VQA improves federated training grounding quality by up to 4.45\%, preserves retained-client utility after deletion, and achieves lower counterfactual gap and prediction disagreement than representative unlearning baselines.
In summary, our contributions are:
\begin{itemize}
    \item We formulate the first federated learning and federated unlearning framework for a chain-of-thought planner in VideoQA, and identify the key challenges of history-dependent deletion, heterogeneous temporal reasoning, and trace-level forgetting quality.

    \item We propose \emph{FedCoT-VQA}, a modular planner-centric design with \textsc{PSP}, \textsc{SSA}, and \textsc{RUM}, which together enable compact federated planner adaptation, deletion-ready aggregation, and efficient client-level unlearning without full retraining.

    \item We conduct experiments on representative VideoQA benchmarks, showing that FedCoT-VQA remains competitive in federated training utility and achieves stronger unlearning and forgetting quality than representative federated unlearning baselines.
\end{itemize}

\begin{figure*}[t]
    \centering
    \subfigure[Standard VideoQA inference pipeline]
    {\includegraphics[width=0.505\textwidth]{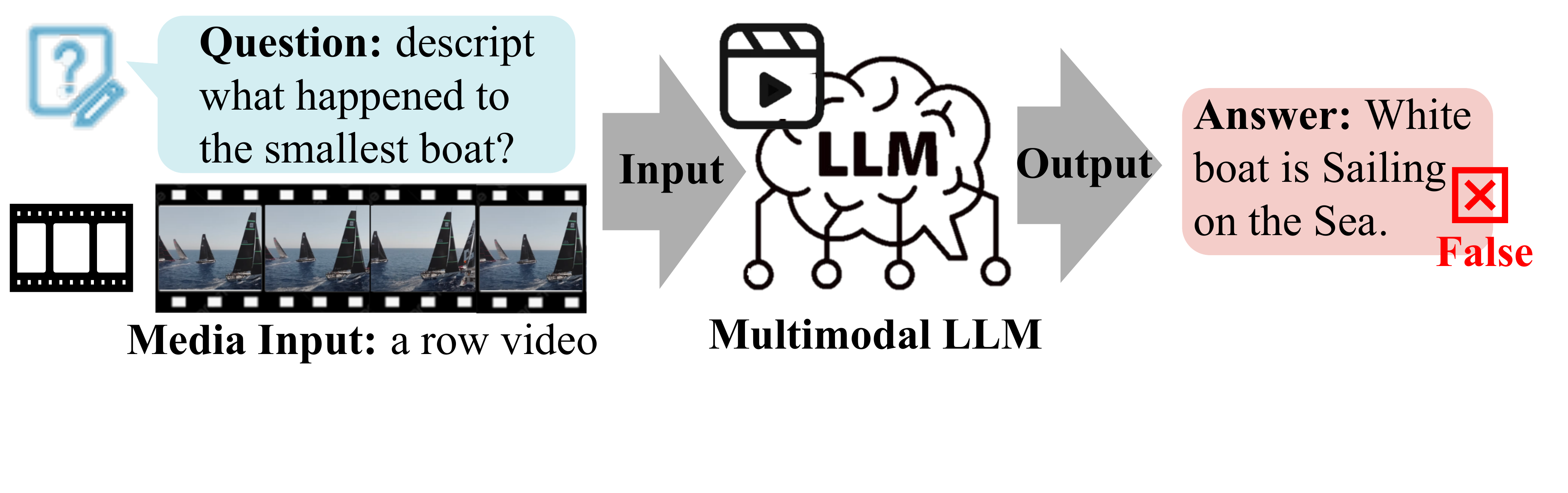}}\label{fig:non_cot_videoqa}
    \hfill
    \subfigure[CoT-enhanced VideoQA pipeline]{
        \includegraphics[width=0.435\textwidth]{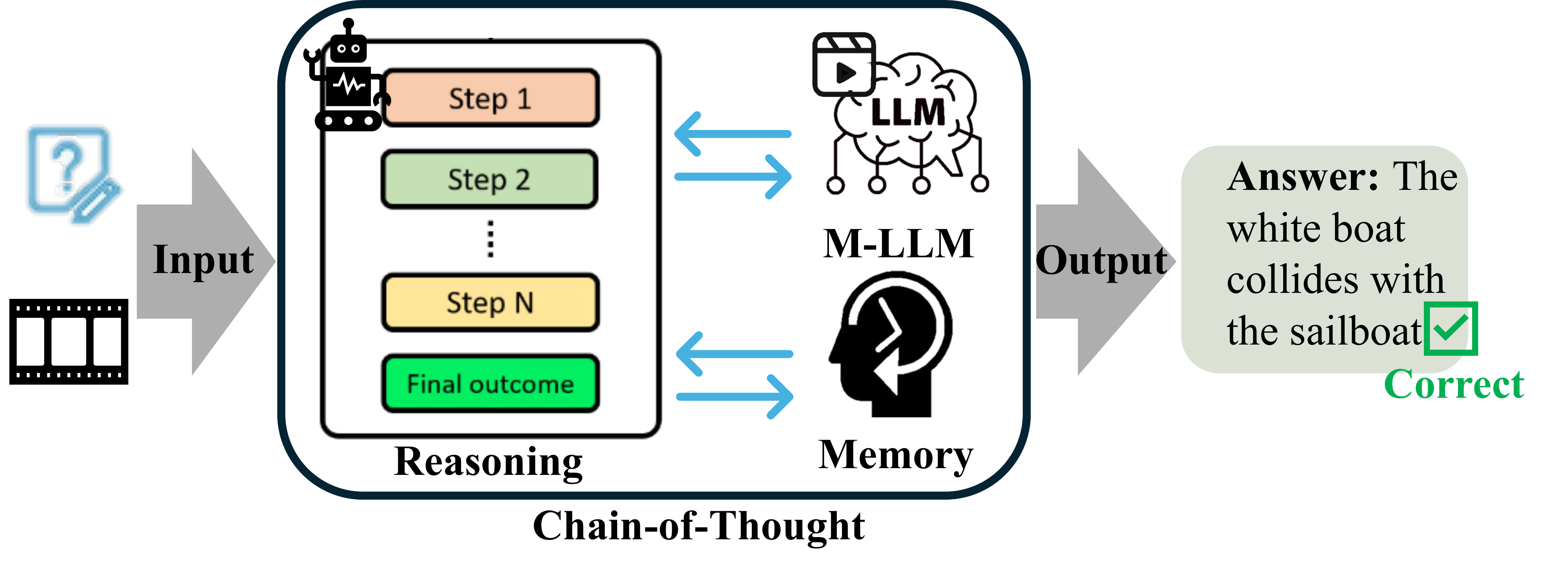}
    } \label{fig:cot_videoqa}
    \caption{Comparison of VideoQA inference paradigms. (a) A direct approach where the Multimodal LLM predicts the answer solely based on the video and question. (b) A Chain-of-Thought framework where a reasoning module orchestrates a multi-step process, leveraging memory and iterative M-LLM interactions to generate the answer.}
    \label{fig:videoqa_pipelines}

\end{figure*}

\section{Background}\label{sec:background}

\subsection{Video CoT for VideoQA}
Video question answering (VideoQA) aims to answer natural language questions about videos and is a core task in video understanding.
Early approaches relied on specialized video encoders and task-specific heads, where temporal reasoning was implemented via hand-designed recurrence, temporal pooling, or attention over pre-extracted visual features.
Recent progress powered by large vision--language models (VLMs) has shifted VideoQA toward general-purpose, instruction-following pipelines that unify perception and language reasoning, enabling broader applications such as \textcolor{black}{video assistants, surveillance querying, and sports analytics.}

However, VideoQA remains challenging.
Many questions require tracking entities and scene state over time, resolving references (who/what/where), and reasoning over temporal relations such as before/after and causality, where relevant evidence is often sparse and temporally dispersed.
Processing the full video at high resolution is computationally expensive, \textcolor{black}{whereas coarse subsampling} can miss short but decisive moments (e.g., a brief interaction or sign), degrading answer quality.
To address this trade-off, recent studies propose modular CoT-style pipelines that separate \emph{evidence planning} from \emph{answer generation} \cite{video_of_thought_2024,frame_voyager_2024,mllm_frame_selection_2025,qframe_2025}.

\textcolor{black}{\textbf{CoT reasoning}} introduces multi-step problem solving by decomposing a query into intermediate steps that structure how evidence is gathered and how conclusions are obtained via a \textit{CoT planner}.
This decomposition aligns naturally with VideoQA as long videos introduce a \emph{where-to-look} problem before a \emph{what-to-answer} problem: relevant evidence is often sparse, temporally dispersed, and only becomes meaningful after combining multiple moments in the correct order.
By making intermediate reasoning steps explicit, CoT guides temporal evidence selection, e.g., choosing which frames to inspect, reducing redundant visual computation, and improving robustness.

As illustrated in Fig.~\ref{fig:videoqa_pipelines} (a), traditional VideoQA frameworks typically employ a direct inference paradigm, where a Multimodal LLM (M-LLM) processes the visual input and the user query in a single pass to generate a response. In contrast, the CoT-enhanced approach (Fig.~\ref{fig:videoqa_pipelines}(b)) incorporates an explicit reasoning module. Rather than generating an immediate answer, this module decomposes the task into sequential steps, iteratively interacting with the M-LLM and utilizing an external memory to store intermediate contexts.
This structured execution enables the system to accumulate evidence and refine its logic before deriving the final answer.

\begin{figure*}[t]
    \centering
    \subfigure[Standard Federated Learning Pipeline]{\includegraphics[width=0.75\columnwidth]{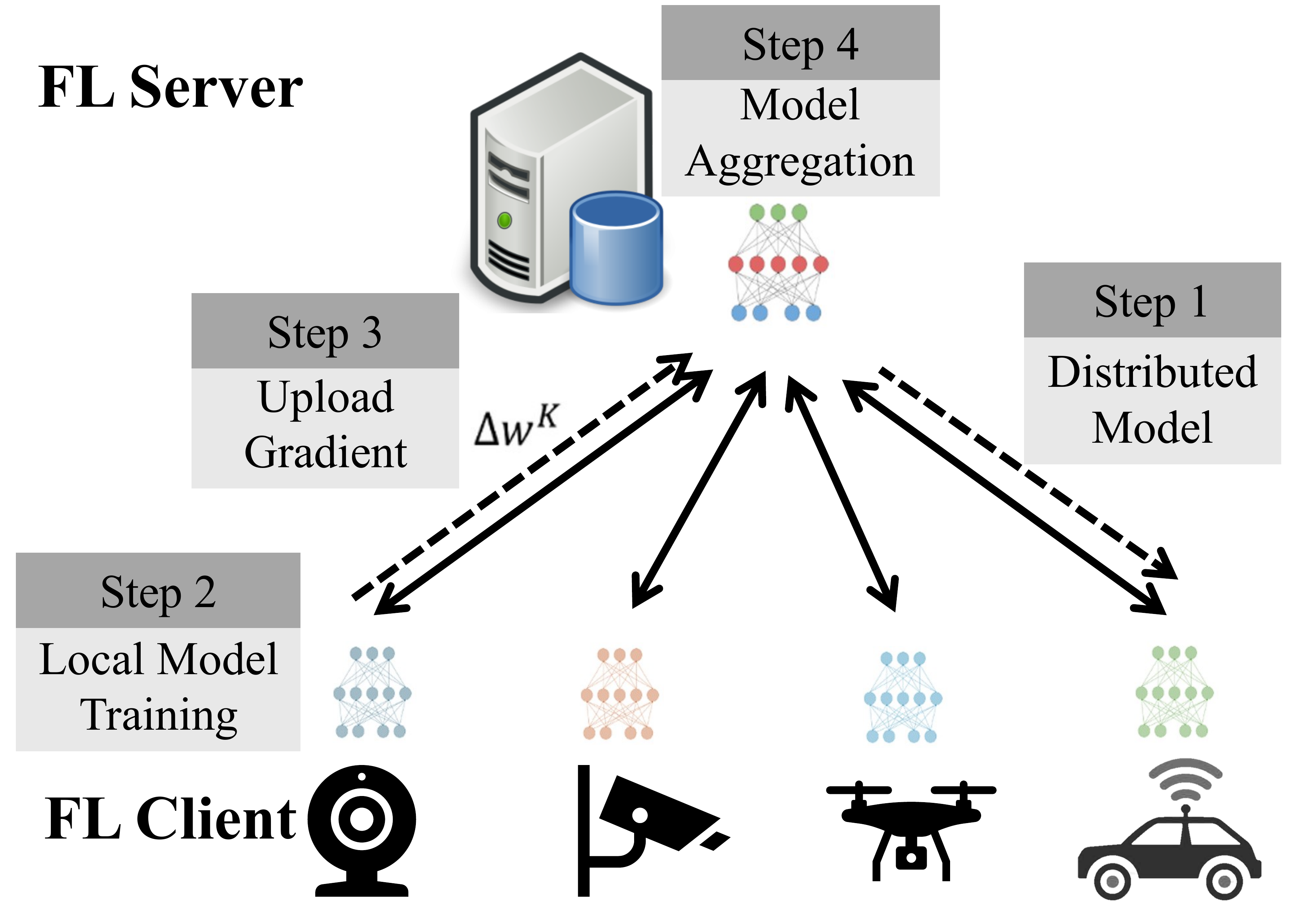}}\label{fig:federated_learning}
    \hspace{35pt}
    \subfigure[Federated Unlearning Mechanism]{
        \includegraphics[width=0.75\columnwidth]{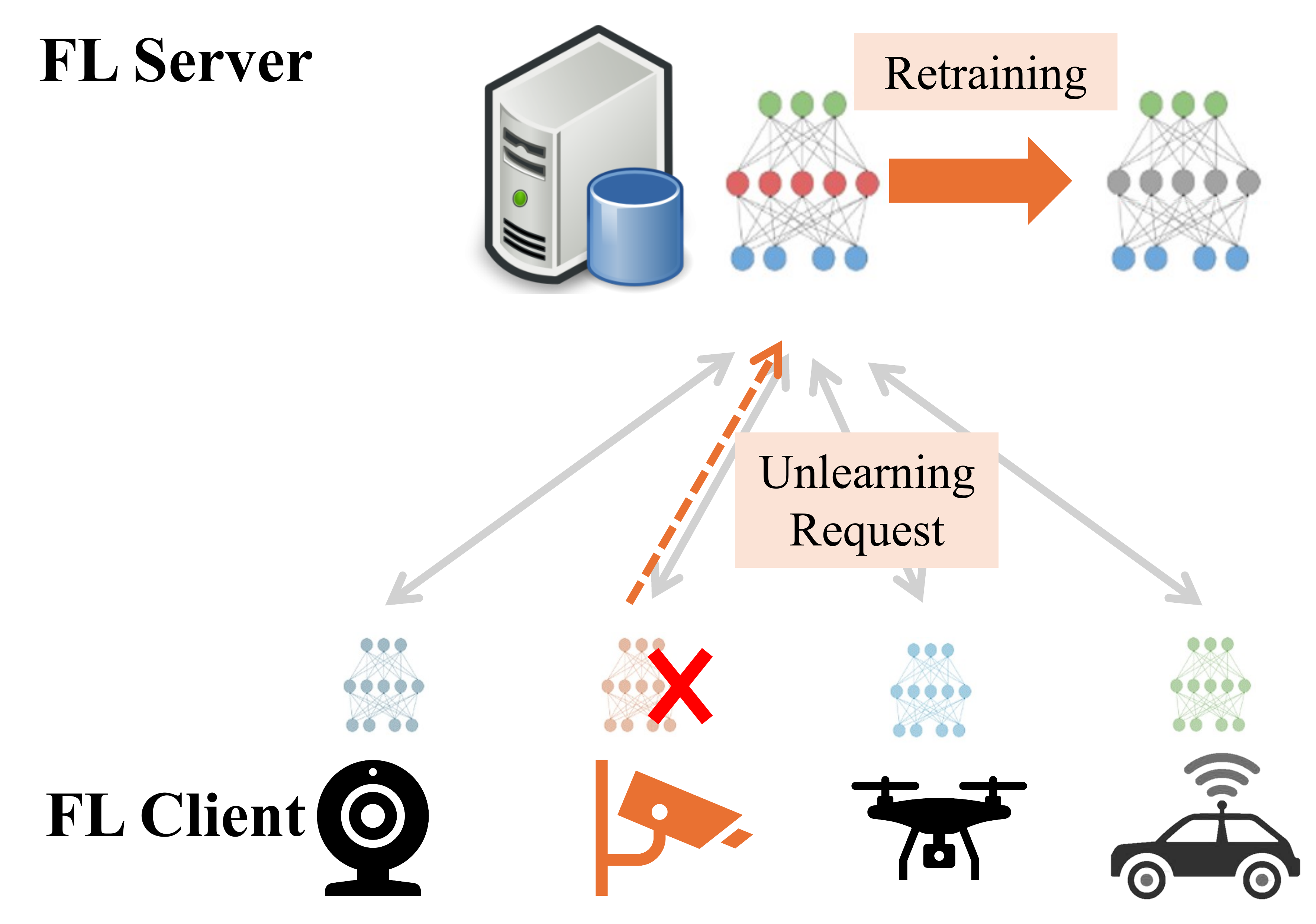}
    }\label{fig:federated_unlearning}
    \caption{Schematic overview of the proposed framework. (a) The standard Federated Learning cycle consists of global model distribution, local training on private data, gradient upload, and server-side aggregation. (b) \textcolor{black}{The Federated Unlearning process, initiated by a client's deletion request, triggers the server} to eliminate the specific client's contribution from the global model (e.g., via retraining or approximate unlearning).}
    \label{fig:fl_framework}

\end{figure*}

Training a CoT planner for VideoQA benefits from diverse data that is naturally distributed across many devices and organizations. In practice, the planner must learn \emph{where to look} in long videos under highly heterogeneous conditions, including different scenes, viewpoints, motion patterns, and question types across deployment sites such as homes, campuses, and hospitals. Such diversity is important for generalization, but centralizing these long videos is often impractical due to bandwidth and storage costs, privacy concerns, and cross-organization data restrictions\textcolor{black}{~\cite{lu2022preva,hu2021feva,lu2026privacy}}. These challenges make federated learning a natural solution for the CoT planner, which is a lightweight and modular component that can be trained collaboratively without retraining the full VideoQA backbone.

Current Video CoT studies mainly focus on accuracy and efficiency, while paying much less attention to how such planners can be trained under federated settings or how client contributions should be managed, especially when deletion requests arise. The modular structure of CoT-based VideoQA is useful not only for efficient inference but also for federated optimization. Instead of federating the entire VideoQA model, we can train only the planner that controls temporal evidence selection across clients.

\subsection{Federated Learning for CoT Planners}
\textit{Federated learning} (FL) provides a natural mechanism for collaborative training of CoT planners for VideoQA when training data is decentralized.
Instead of uploading raw videos to a central server, each participant updates the local planner using its own videos and queries, and only model updates, e.g., gradients or parameter deltas, are shared for aggregation into a global planner, as illustrated in Fig.~\ref{fig:federated_learning}.
This \emph{data-local, update-aggregate} workflow enables learning from a wide range of environments and user intents while keeping the training pipeline lightweight.  The global coordination focuses on the modular planner, and the large VideoQA backbone can remain fixed without frequent updates.

FL is naturally suitable for CoT VideoQA for two reasons.
First, the planner is much smaller than the full VideoQA backbone, so local training and communication remain practical even when videos are long and heterogeneous.
Second, the planner directly models the \emph{where-to-look} decision, which is exactly the component that must adapt across sites with different environments, camera views, and query distributions.

\textcolor{black}{Privacy risks also arise from intermediate representations that can reveal private inputs~\cite{lu2026p4llm}.} However, training CoT planners also introduces privacy risks beyond raw data exposure.
Planner outputs and intermediate traces can encode sensitive behavioral signals, such as client-specific viewing patterns, recurring scene cues, and \textcolor{black}{rare temporal combinations that uniquely identify clients.}
Moreover, these signals may be reflected in shared model updates, even when the underlying videos never leave the client.

At the same time, FL alone does not satisfy the deletion and compliance requirements that arise in real deployments.
Video data ownership and access rights could change over time.
For example, participants may terminate contracts, revoke data-sharing consent, or \textcolor{black}{exercise their right to be forgotten}, requiring that their past training influence be removed from the shared model.
In a CoT-style VideoQA setting, this requirement is particularly critical as the trained planner captures \emph{where-to-look} behavior that can reflect client-specific environments and reveal identifiable client patterns.

\subsection{Federated Unlearning for CoT Planners}
\emph{Federated unlearning} (FU) formalizes this requirement by updating the trained global planner to approximate the \emph{counterfactual} model that would have been obtained had a specific client never participated (Fig.~\ref{fig:federated_unlearning}).
In standard federated optimization, this goal is already difficult because client influence is history dependent: it is shaped by round-wise client sampling, repeated aggregation, and optimizer state. As a result, deleting one client cannot, in general, be achieved by simply subtracting one update from the final model.

This challenge becomes sharper for CoT planners for VideoQA.
Exact compliance via full retraining from scratch remains expensive, even if only the planner is trained, due to long training horizons and repeated communication.
Therefore, FU for CoT planners must fit both the optimization structure of FL and the behavioral structure of VideoQA reasoning.
These practical constraints motivate FU mechanisms that are both efficient and effective for CoT planners, providing strong deletion while preserving the utility of evidence planning and downstream VideoQA accuracy.

\section{Problem Formulation}\label{sec:pro}
\subsection{VideoQA with a CoT Planner}\label{subsec:videoqa_cot}
Let $\mathcal{D}=\{(v_n,q_n,y_n)\}_{n=1}^{N}$ denote a VideoQA dataset, where $v_n$ is a video, $q_n$ is a natural-language question, and $y_n$ is the corresponding answer.
We formulate VideoQA with three functional components: a \textit{planner}, an \textit{extractor}, and an \textit{answerer}.
Specifically, the planner $\pi_{\phi}$ predicts a question-conditioned plan $p_n$ over the input video. The extractor $\mathcal{E}$ maps this plan to concrete visual evidence $e_n$, and the answerer $f_{\theta}$ infers the final answer from the extracted evidence $e_n$ and the input question $q_n$. It is written as:
\[
p_n=\pi_{\phi}(v_n,q_n),\quad
e_n=\mathcal{E}(v_n,p_n),\quad
\hat{y}_n=f_{\theta}(e_n,q_n,p_n).
\]

In a standard non-CoT planner, $p_n$ may simply represent compact evidence-selection decisions, such as relevant clips, temporal spans, or frame indices.
In this paper, we focus on the CoT planner, which outputs \textit{a structured sequential reasoning trace} $c_n$ over $v_n$ rather than a typical flat selection result $p_n$.
Formally, for sample $n$, the generic plan variable $p_n$ takes the form of a CoT intermediate reasoning trajectory:
\begin{equation} \label{eq:cot_trace}
p_n \equiv c_n=\pi_{\phi}(v_n,q_n)=\big(z_{n,1},z_{n,2},\ldots,z_{n,K_n}\big),
\end{equation}
where $K_n$ is the number of reasoning steps for sample $n$ and each reasoning step
\begin{equation}\label{eq:reason_step}
z_{n,k}=(s_{n,k},t_{n,k},r_{n,k}),\quad \textcolor{black}{k=1,2,\ldots,K_n}
\end{equation}
contains a \textit{temporal span} $(s_{n,k},t_{n,k})$ and a \textit{reasoning state} $r_{n,k}$.
Here, the temporal span specifies \emph{where} to inspect in the video, i.e., $s_{n,k}$ and $t_{n,k}$ denote the start and end timestamps of the video segment selected at step $k$ for sample $n$.
$r_{n,k}$ specifies \emph{why} that segment is relevant, e.g., the semantic role or sub-goal of the $k$-th reasoning step.

Overall, the CoT trace $c_n$ specifies both the evidence to inspect via $(s_{n,k},t_{n,k})$ and the step-wise reasoning structure in Eq.~\ref{eq:cot_trace}.
The reasoning state $r_{n,k}$ provides step-level guidance for downstream evidence alignment and answer generation.

\noindent\textbf{CoT-conditioned extraction and answering.}
The extractor $\mathcal{E}$ is conditioned on each reasoning step instead of only on the whole question, preserving alignment between reasoning and visual evidence as:
\[
e_{n,k}=\mathcal{E}(v_n,z_{n,k}),
\qquad
e_n=\big(e_{n,1},\ldots,e_{n,K_n}\big).
\]

The answerer $f_{\theta}$ then produces the answer $\hat{y}_n$ by aggregating the evidence sequence within the VideoQA pipeline:
\[
\hat{y}_n=f_{\theta}\!\left(q_n,\, e_n,\, c_n\right)=f_{\theta}\!\left(q_n,\{(e_{n,k},r_{n,k})\}_{k=1}^{K_n}\right).
\]
Here, the CoT planner shapes both \emph{what evidence is consulted}, i.e., the evidence path $e_n$, and \emph{how the answer is constructed}, i.e., the reasoning path $\{r_{n,k}\}_{k=1}^{K_n}$.
That is, the system first predicts a CoT reasoning trace $c_n$, \textcolor{black}{then aligns each reasoning step with visual evidence $e_n$}, and finally generates the answer $\hat{y}_n$ using both the grounded evidence and the intermediate reasoning trajectory.

\noindent\textbf{\textcolor{black}{Learning objective for the CoT planner.}}
Before introducing the federated setting, we first formulate the learning objective for the CoT planner $\pi_{\phi}$.
We train $\phi$ by
\begin{equation}
\label{eq:central_obj}
\begin{aligned}
\min_{\phi}\ \mathcal{L}_{\mathrm{train}}
&=
\frac{1}{N}\sum_{n=1}^{N}
\Big[
\ell_{\mathrm{qa}}(\hat{y}_n,y_n)
+\lambda_1\ell_{\mathrm{budget}}(c_n) \\
&
+\lambda_2\ell_{\mathrm{cover}}(c_n;v_n,q_n)
+\lambda_3\ell_{\mathrm{stable}}(c_n)
\Big].
\end{aligned}
\end{equation}
Here, $\ell_{\mathrm{qa}}$ measures the answer prediction error, as is standard in VideoQA learning~\cite{lei2020tvqaplus,li2022igv}.
$\ell_{\mathrm{budget}}$ penalizes excessive evidence usage, e.g., \textcolor{black}{too many reasoning steps} or overly long selected spans~\cite{yao2025gens,xu2026revise}.
$\ell_{\mathrm{cover}}$ encourages the trace to cover temporally relevant evidence for answering the question~\cite{lei2020tvqaplus,wang2021wstan,li2022igv}.
$\ell_{\mathrm{stable}}$ regularizes incoherent transitions across adjacent reasoning steps~\cite{wang2021wstan,zhao2020bottomup}.

\subsection{Federated Training of the CoT Planner}\label{subsec:fl_cot_planner}
We now consider a federated setting in which the training data are distributed across multiple clients and cannot be directly pooled at a central server, requiring the planner to be trained collaboratively across clients rather than centrally.
We consider a federated setting with $M$ clients, where the global VideoQA dataset is partitioned as
\[
\mathcal{D}=\bigcup_{m=1}^{M}\mathcal{D}_m,\qquad
\mathcal{D}_i\cap\mathcal{D}_j=\emptyset,\ \forall i\neq j.
\]
Here, $\mathcal{D}_m=\{(v_{m,n},q_{m,n},y_{m,n})\}_{n=1}^{N_m}$ denotes the local dataset held by client $m$, and $N_m=|\mathcal{D}_m|$ is the number of local samples. The total number of samples is $N=\sum_{m=1}^{M}N_m$. \textcolor{black}{Since client data are generated} from different users, devices, or deployment environments, the local data distributions may be heterogeneous and non-IID.

To reduce communication and preserve the reasoning capability of the large pretrained VideoQA backbone, the extractor $\mathcal{E}$ and the answerer $f_{\theta}$ remain fixed,
and \textcolor{black}{the full CoT planner parameter set can be further decomposed as}
\begin{equation}\label{eq:fed_split}
    \phi_{\mathrm{full}}=\phi_{\mathrm{tr}}\cup\phi_{\mathrm{frz}},\qquad
\phi_{\mathrm{tr}}\cap\phi_{\mathrm{frz}}=\emptyset,
\end{equation}
where $\phi_{\mathrm{tr}}$ denotes the trainable parameter subset exposed to federated optimization, and $\phi_{\mathrm{frz}}$ denotes the remaining frozen planner parameters.
Throughout federated training, the server broadcasts and aggregates only $\phi_{\mathrm{tr}}$.

\noindent\textbf{Client local objective.}
Given the above parameter split, the client-local training objective is to minimize:
\begin{equation}\label{eq:fl_ov_loss}
\resizebox{0.905\linewidth}{!}{$
\begin{aligned}
\mathcal{L}_m(\phi_{\mathrm{tr}};\phi_{\mathrm{frz}})
&= \frac{1}{N_m}\sum_{n=1}^{N_m} \Big[
\ell_{\mathrm{qa}}(\hat{y}_{m,n},y_{m,n})
+ \lambda_1 \ell_{\mathrm{budget}}(c_{m,n}) \\
&
+ \lambda_2 \ell_{\mathrm{cover}}(c_{m,n};v_{m,n},q_{m,n})
+ \lambda_3 \ell_{\mathrm{stable}}(c_{m,n})
\Big].
\end{aligned}
$}
\end{equation}
Similar to the centralized objective in Eq.~\ref{eq:central_obj}, this local loss is evaluated only over the client-local dataset $\mathcal{D}_m$.

\noindent\textbf{Server global objective.}
The goal of federated training is to learn a global planner parameter $\phi_{\mathrm{tr}}$ that minimizes the weighted sum of local objectives:
\begin{equation} \label{eq:client_fl_loss}
\min_{\phi_{\mathrm{tr}}}\ \mathcal{L}_{\mathrm{FL}}(\phi_{\mathrm{tr}};\phi_{\mathrm{frz}})
=
\sum_{m=1}^{M}\frac{N_m}{N}\,\mathcal{L}_m(\phi_{\mathrm{tr}};\phi_{\mathrm{frz}}).
\end{equation}

This objective aggregates the client-local losses into a global federated objective that serves as the decentralized counterpart of the centralized training objective.

It seeks a single global CoT planner that performs well across decentralized client distributions \textcolor{black}{without requiring raw training data to leave local devices}.

\noindent\textbf{Round-based training process.}
Federated training proceeds in several communication rounds. At round $t$, the server broadcasts the current global planner parameter $\phi_{\mathrm{tr}}^{(t)}$ to participating clients $\mathcal{S}^{(t)}$. Each client performs several steps of local optimization on $\mathcal{L}_m(\phi_{\mathrm{tr}}^{(t)};\phi_{\mathrm{frz}})$ and returns an updated planner parameter $\phi_m^{(t+1)}$.
The server then aggregates the received local planners to obtain the next global planner, as follows:
\[
\phi_{\mathrm{tr}}^{(t+1)}
=
\sum_{m\in\mathcal{S}^{(t)}}
\frac{N_m}{\sum_{j\in\mathcal{S}^{(t)}}N_j}
\phi_m^{(t+1)}.
\]

Here, federated training optimizes the CoT planner collaboratively over distributed client data without centralizing raw videos.

\subsection{Federated Unlearning of the CoT Planner}\label{subsec:fu_cot_planner}

After federated training, some clients may request the removal of their contributions from the global planner.
Let $\mathcal{U}\subseteq\{1,\ldots, M\}$ denote the set of clients to be unlearned, and let the union of their local datasets $\mathcal{D}_{\mathcal{U}}$ and its complement $\mathcal{D}_{-\mathcal{U}}$ be defined as
\[
\mathcal{D}_{\mathcal{U}}=\bigcup_{m\in\mathcal{U}}\mathcal{D}_m, \qquad
\mathcal{D}_{-\mathcal{U}}=\bigcup_{m\notin\mathcal{U}}\mathcal{D}_m.
\]

\begin{figure*}[t!]
    \centering
    \includegraphics[width=0.95\textwidth]{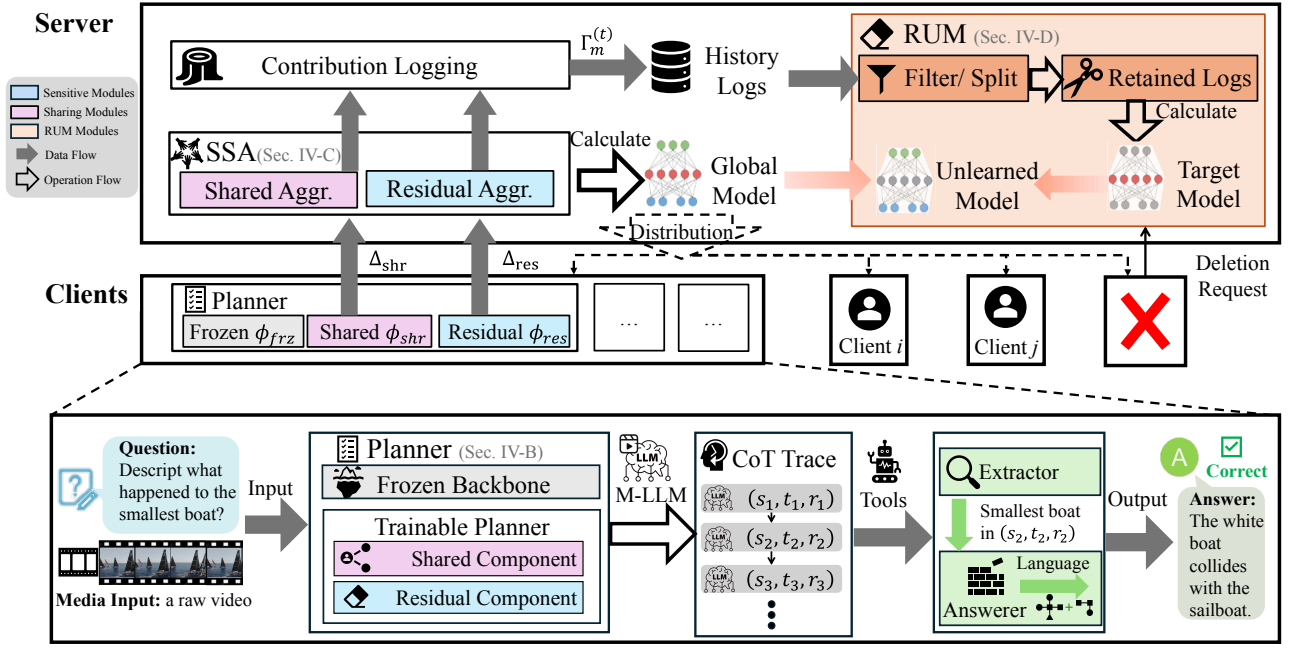}
    \caption{The overall architecture of our proposed framework. The system consists of three main modules: planner-side partitioning (PSP), server-side aggregation (SSA), and the residual unlearning module (RUM). The blue arrows indicate the forward data flow, while the red dashed lines denote the update flow during training and unlearning.}
    \label{fig:FedCoT_framework}
\end{figure*}

\textcolor{black}{Since the full planner parameter set is decomposed as} $\phi_{\mathrm{full}}=\phi_{\mathrm{tr}}\cup\phi_{\mathrm{frz}}$, where only the trainable subset $\phi_{\mathrm{tr}}$ participates in federated optimization, the goal of federated unlearning is to remove the influence of $\mathcal{D}_{\mathcal{U}}$ from the trained trainable planner parameter $\phi_{\mathrm{tr}}$ while keeping the frozen planner parameter $\phi_{\mathrm{frz}}$ unchanged and preserving the model utility.

Ideally, the updated trainable planner should match the counterfactual planner that would have been obtained had the clients in $\mathcal{U}$ never participated in federated training.
Formally, let $\phi_{\mathrm{tr}}^{-{\mathcal{U}}}$ denote the ideal counterfactual solution defined by
\[
\phi_{\mathrm{tr}}^{-{\mathcal{U}}}
=
\arg\min_{\phi_{\mathrm{tr}}}\ \mathcal{L}_{\mathrm{FL}}^{-{\mathcal{U}}}(\phi_{\mathrm{tr}};\phi_{\mathrm{frz}}),
\]
\[
\mathcal{L}_{\mathrm{FL}}^{-{\mathcal{U}}}(\phi_{\mathrm{tr}};\phi_{\mathrm{frz}})
=
\sum_{m\notin\mathcal{U}}
\frac{N_m}{N_{-\mathcal{U}}}\,\mathcal{L}_m(\phi_{\mathrm{tr}};\phi_{\mathrm{frz}}),~N_{-\mathcal{U}}=\sum_{m\notin\mathcal{U}} N_m.
\]
Here, $\mathcal{L}_m(\phi_{\mathrm{tr}};\phi_{\mathrm{frz}})$ is the local CoT-planner loss in Eq.~\ref{eq:fl_ov_loss}, and the above aggregation excludes all deleted clients.

\noindent\textbf{Unlearning objectives.}
In practice, fully retraining the federated CoT planner on $\mathcal{D}_{-\mathcal{U}}$ is computationally expensive.
Therefore, the federated unlearning problem is to find an efficient updated planner $\phi_{\mathrm{tr}} \rightarrow \tilde{\phi_{\mathrm{tr}}}^{-{\mathcal{U}}}$.
Here, $\tilde{\phi_{\mathrm{tr}}}^{-{\mathcal{U}}}$ approximates the counterfactual planner $\phi_{\mathrm{tr}}^{-{\mathcal{U}}}$, removes the contribution of deleted clients, and preserves VideoQA utility on the remaining clients.
After unlearning, the original trainable planner parameter $\phi_{\mathrm{tr}}$ is replaced by the unlearned one $\tilde{\phi_{\mathrm{tr}}}^{-{\mathcal{U}}}$.

\textcolor{black}{Unlearning has two quality objectives.} First, the updated planner should be close to the counterfactual retrained model, so that the deleted clients no longer affect the CoT reasoning traces or the final VideoQA outputs. Second, it should still perform well on the retained clients, i.e., its answer accuracy and reasoning quality on $\mathcal{D}_{-\mathcal{U}}$ should remain close to those of the ideal counterfactual model.

\section{FedCoT-VQA Design}\label{sec:design}
\subsection{Overview}\label{subsec:overview}
FedCoT-VQA is a modular framework for federated training and unlearning of a CoT planner in VideoQA. It consists of three tightly coupled modules: \textit{PSP}, which exposes a compact planner-side adaptation space for federated client training; \textit{SSA}, which aggregates client updates and records a compact server-side contribution history; and \textit{RUM}, which removes deleted-client influence without full retraining. Together, these three modules form an end-to-end pipeline for federated learning and unlearning of the CoT planner.

Fig.~\ref{fig:FedCoT_framework} illustrates the overall workflow of FedCoT-VQA. At communication round $t$, the server broadcasts the current trainable planner parameters $(\phi_{\mathrm{shr}}^{(t)},\phi_{\mathrm{res}}^{(t)})$ to the selected clients, while keeping the frozen planner parameter $\phi_{\mathrm{frz}}$ unchanged.
On each client, \textsc{PSP} attaches to the underlying CoT planner and restricts local adaptation to a shared component $\phi_{\mathrm{shr}}$ and a residual component $\phi_{\mathrm{res}}$. The client then performs local CoT-based VideoQA training and returns the corresponding updates $(\Delta_{\mathrm{shr},m}^{(t)},\Delta_{\mathrm{res},m}^{(t)})$. The server-side \textsc{SSA} aggregates these updates into the next global planner and records a compact contribution log $\Gamma_m^{(t)}$ for later unlearning. When a deletion request arrives, \textsc{RUM} reuses this stored history to reconstruct a retained-only replay trajectory and then revises the more deletion-sensitive part of the planner, thereby producing an unlearned planner that approximates the counterfactual model without full retraining.

\textcolor{black}{The design of FedCoT-VQA addresses three challenges}:
\begin{itemize}

\item \textbf{Challenge 1: How to support heterogeneous client adaptation without losing a compact and deletion-ready CoT planner? }Even after restricting optimization to a compact trainable subset of the planner, different clients may still exhibit different video domains, question distributions, and reasoning styles. Treating all trainable parameters uniformly can therefore limit adaptation flexibility and make later client-level deletion less precise.
\textcolor{black}{FedCoT-VQA addresses this challenge through} \textsc{PSP}, which further organizes the trainable planner parameters into shared and residual parts on top of the frozen backbone. It preserves compact federated optimization while improving heterogeneity adaptation and later unlearning readiness.

\item\textbf{Challenge 2: How to aggregate heterogeneous client updates while preserving training stability and future deletability?}
Different clients may have different video domains, question distributions, and reasoning styles, so directly averaging all planner updates weakens stability and client-level deletion.
\textcolor{black}{FedCoT-VQA addresses this challenge through} \textsc{SSA}, which separately aggregates the shared and residual updates and maintains a compact server-side contribution log. It prepares the information for later unlearning while keeping federated training stable under client heterogeneity.

\item \textbf{Challenge 3: How to remove deleted-client influence efficiently without full retraining?}
Simply subtracting deleted-client updates is often insufficient for FU. FedCoT-VQA designs \textsc{RUM}, which combines retained-update replay with \textcolor{black}{targeted revision of the more deletion-sensitive part} of the planner. As a result, deleted-client influence can be removed more precisely while preserving VideoQA reasoning utility.
\end{itemize}

Overall, FedCoT-VQA follows a planner-centric design, where \textsc{PSP} makes an existing CoT planner trainable under federated constraints, \textsc{SSA} makes client updates aggregatable and deletion-ready, and \textsc{RUM} makes the resulting planner efficiently unlearnable. \reviewblue{This modular pipeline enables the framework to jointly balance training efficiency, reasoning quality, and client-level deletability in CoT-based VideoQA.}

\subsection{Planner-Side Partitioning for Federated Training (PSP)}\label{subsec:trace_adapter}

FedCoT-VQA first introduces \textsc{PSP}, a \textbf{p}lanner-\textbf{s}ide \textbf{p}artitioning mechanism that can be attached to an existing CoT planner to enable federated training and unlearning.
The key idea is to preserve the original planner's reasoning capability, while exposing a compact and controllable adaptation space for federated optimization and later deletion.

As introduced in Eq.~\ref{eq:fed_split}, the full planner parameter set is decomposed into a trainable part and a frozen part.
In \textsc{PSP}, we further refine this decomposition by organizing the trainable part into two cooperative components: a shared adaptation component and a residual adaptation component.
Specifically, let $\pi_{\phi_{\mathrm{full}}}$ denote an existing CoT planner for VideoQA, and decompose
\[
\phi_{\mathrm{full}}=\phi_{\mathrm{shr}}\cup\phi_{\mathrm{res}}\cup\phi_{\mathrm{frz}},
\]
\[
\phi_{\mathrm{shr}}\cap\phi_{\mathrm{res}}
=
\phi_{\mathrm{shr}}\cap\phi_{\mathrm{frz}}
=
\phi_{\mathrm{res}}\cap\phi_{\mathrm{frz}}
=
\emptyset.
\]

Here, $\phi_{\mathrm{frz}}$ denotes the frozen planner parameters, $\phi_{\mathrm{shr}}$ denotes the trainable component that captures shared task-level reasoning knowledge across clients, and $\phi_{\mathrm{res}}$ denotes a lightweight residual adaptation component that absorbs client-specific reasoning preferences and is therefore suitable for later removal during unlearning.
In this way, \textsc{PSP} does not change the basic structure of the CoT planner, but partitions its parameter space into frozen, shared, and deletion-sensitive components for federated learning and unlearning.

\begin{algorithm2e}[t]
\caption{Score-Guided Planner Partition}
\label{alg:psp_partition}
\small
\KwIn{Pretrained CoT planner $\pi_{\phi_{\mathrm{full}}}$; candidate component set $\{\omega_j\}$ covering $\phi_{\mathrm{full}}$; dataset $\mathcal{D}$; weights $\alpha,\beta,\gamma$; thresholds $\tau_{\mathrm{frz}},\tau_{\mathrm{del}}$}
\KwOut{Frozen subset $\phi_{\mathrm{frz}}$, shared trainable subset $\phi_{\mathrm{shr}}$, residual trainable subset $\phi_{\mathrm{res}}$}

Initialize $\phi_{\mathrm{frz}}\leftarrow\emptyset$, $\phi_{\mathrm{shr}}\leftarrow\emptyset$, and $\phi_{\mathrm{res}}\leftarrow\emptyset$\;

\ForEach{planner component $\omega_j \in \{\omega_j\}$}{
    Compute $S^{\mathrm{util}}_j$ according to Eq.~\ref{eq:psp_util_score}\;

    Compute $S^{\mathrm{cot}}_j$ according to Eq.~\ref{eq:psp_cot_score}\;

    Compute $S^{\mathrm{del}}_j$ according to Eq.~\ref{eq:psp_del_score}\;

    Compute the joint score $S_j$ according to Eq.~\ref{eq:psp_joint_score}\;

    \uIf{$S_j < \tau_{\mathrm{frz}}$}{
        $\phi_{\mathrm{frz}}\leftarrow \phi_{\mathrm{frz}}\cup\{\omega_j\}$\;
    }
    \uElseIf{$S^{\mathrm{del}}_j \ge \tau_{\mathrm{del}}$}{
        $\phi_{\mathrm{res}}\leftarrow \phi_{\mathrm{res}}\cup\{\omega_j\}$\;
    }
    \Else{
        $\phi_{\mathrm{shr}}\leftarrow \phi_{\mathrm{shr}}\cup\{\omega_j\}$\;
    }
}
\Return{$\phi_{\mathrm{frz}},\phi_{\mathrm{shr}},\phi_{\mathrm{res}}$}\;
\end{algorithm2e}

\noindent\textbf{Score-guided parameter partition.}
Rather than splitting planner parameters heuristically, \textsc{PSP} assigns each candidate planner component $\omega_j$ from a predefined collection $\{\omega_j\}$ covering the planner parameter space $\phi_{\mathrm{full}}$ a joint importance score that reflects its contribution to VideoQA utility, CoT-trace quality, and deletion sensitivity.
\textcolor{black}{This scoring approach is inspired by importance-aware and sensitivity-aware parameter-efficient tuning, as used in} AdaLoRA~\cite{zhang2023adalora}, FibecFed~\cite{liu2024fibecfed}, and FUSED~\cite{zhong2025fused}.
Accordingly, we define
\begin{equation}\label{eq:psp_joint_score}
S_j=\alpha S^{\mathrm{util}}_j+\beta S^{\mathrm{cot}}_j+\gamma S^{\mathrm{del}}_j, \qquad \alpha,\beta,\gamma\ge 0,
\end{equation}
where $\{\alpha,\beta,\gamma\}$ control the trade-off among answer quality, reasoning quality, and unlearning-friendliness.
Specifically,

\noindent i) $S^{\mathrm{util}}_j$ measures how strongly $\omega_j$ contributes to answer correctness following the idea of gradient- or importance-based budget allocation in \cite{zhang2023adalora}:
\begin{equation}\label{eq:psp_util_score}
S^{\mathrm{util}}_j
=
\mathbb{E}_{(v,q,y)\sim\mathcal{D}}
\big[
\left\|
\nabla_{\omega_j}\ell_{\mathrm{qa}}(\hat{y},y)
\right\|_2
\big].
\end{equation}

\noindent ii) $S^{\mathrm{cot}}_j$ measures how strongly $\omega_j$ influences the quality of the generated reasoning trace and thus encourages the selected trainable components to preserve reasoning quality:
\begin{equation}\label{eq:psp_cot_score}
S^{\mathrm{cot}}_j
=
\mathbb{E}_{(v,q)\sim\mathcal{D}}
\left[
\left\|
\nabla_{\omega_j}
\big(
\ell_{\mathrm{cover}}(c;v,q)+\ell_{\mathrm{stable}}(c)
\big)
\right\|_2
\right].
\end{equation}

\noindent iii) $S^{\mathrm{del}}_j$ measures how strongly $\omega_j$ captures client-specific variation relevant to unlearning:
\begin{equation}\label{eq:psp_del_score}
S^{\mathrm{del}}_j = \mathrm{Var}_{m} \!\left(\Delta_{m,j}\right),
\end{equation}
where $\Delta_{m,j}$ denotes the update of $\omega_j$ contributed by client $m$.
Intuitively, a component with high cross-client update variance is more likely to encode client-specific reasoning style and is therefore a natural target for deletion-aware adaptation, as also suggested by selective federated unlearning~\cite{zhong2025fused}.

\textcolor{black}{Given the score set} $\{S_j\}$, \textsc{PSP} partitions the planner components into three groups in Algorithm~\ref{alg:psp_partition}.
Components with a low overall score $S_j$ are assigned to the frozen subset $\phi_{\mathrm{frz}}$.
Among the remaining components, \textcolor{black}{those with a high overall score $S_j$} but relatively low deletion sensitivity $S^{\mathrm{del}}_j$ are assigned to $\phi_{\mathrm{shr}}$,
while those with high deletion sensitivity are assigned to $\phi_{\mathrm{res}}$:
\[
\phi_{\mathrm{frz}}=\{\omega_j: S_j<\tau_{\mathrm{frz}}\},
\]
\[
\phi_{\mathrm{res}}=\{\omega_j: S_j\ge\tau_{\mathrm{frz}},\; S^{\mathrm{del}}_j\ge\tau_{\mathrm{del}}\},
\]
\[
\phi_{\mathrm{shr}}=\{\omega_j: S_j\ge\tau_{\mathrm{frz}},\; S^{\mathrm{del}}_j<\tau_{\mathrm{del}}\},
\]
where $\tau_{\mathrm{frz}}$ and $\tau_{\mathrm{del}}$ are partition thresholds.

\begin{algorithm2e}[t]
\caption{Parameter-Efficient Client Adaptation}
\label{alg:psp_local_train}
\small
\KwIn{Client $m$ with local dataset $\mathcal{D}_m=\{(v_{m,n},q_{m,n},y_{m,n})\}_{n=1}^{N_m}$; partitioned planner $\pi_{\phi_{\mathrm{shr}},\phi_{\mathrm{res}},\phi_{\mathrm{frz}}}$; extractor $\mathcal{E}$; answerer $f_\theta$; loss weights $\lambda_1,\lambda_2,\lambda_3$}
\KwOut{Updated trainable planner parameters $\phi_{\mathrm{shr}},\phi_{\mathrm{res}}$ and local update record $\Delta_m$}
Freeze $\phi_{\mathrm{frz}}$, $\mathcal{E}$, and $f_\theta$\;
\For{each local training step}{
    Sample a mini-batch $\mathcal{B}_m \subseteq \mathcal{D}_m$\;

    \ForEach{$(v_{m,n},q_{m,n},y_{m,n}) \in \mathcal{B}_m$}{
        Generate the CoT trace $c_{m,n}$ according to Eq.~\ref{eq:psp_trace_generation}\;

        Extract evidence $e_{m,n}$ according to Eq.~\ref{eq:psp_evidence_extraction}\;

        Predict the answer $\hat{y}_{m,n}$ according to Eq.~\ref{eq:psp_answer_prediction}\;
    }

    Compute the client objective $\mathcal{L}_m$ according to Eq.~\ref{eq:PSP_client_loss}\;

    Update only $\phi_{\mathrm{shr}}$ and $\phi_{\mathrm{res}}$ by minimizing $\mathcal{L}_m$\;
}
$\Delta_m \leftarrow\big(\Delta_m^{\mathrm{shr}},\Delta_m^{\mathrm{res}}\big)$ \;
Record the local trainable-parameter update $\Delta_m$ \;

\Return{updated $\phi_{\mathrm{shr}},\phi_{\mathrm{res}}$ and $\Delta_m$}\;
\end{algorithm2e}

\noindent\textbf{Parameter-efficient local adaptation.}
Since only the trainable subsets $\phi_{\mathrm{shr}}$ and $\phi_{\mathrm{res}}$ are optimized, each client performs local adaptation in a compact planner-side parameter space. \reviewblue{This partitioning design has three advantages.} First, it greatly reduces communication and computation compared with federating the full VideoQA model. Second, it limits cross-client interference to a small and well-defined adaptation space. Third, it localizes client influence to the trainable planner-side updates, which later enables efficient server-side aggregation and client-level unlearning.

\noindent\textbf{Local training under the partitioned planner.}
On client $m$, given a local sample $(v_{m,n},q_{m,n},y_{m,n})$, the partitioned planner first produces
\begin{equation} \label{eq:psp_trace_generation}
c_{m,n}=\pi_{\phi_{\mathrm{shr}},\phi_{\mathrm{res}},\phi_{\mathrm{frz}}}(v_{m,n},q_{m,n}),
\end{equation}
\begin{equation}\label{eq:psp_evidence_extraction}
e_{m,n}=\big(\mathcal{E}(v_{m,n},z_{m,n,1}),\ldots,\mathcal{E}(v_{m,n},z_{m,n,K_{m,n}})\big),
\end{equation}
and then the answerer $f_{\theta}$ predicts
\begin{equation}\label{eq:psp_answer_prediction}
\hat{y}_{m,n}=f_{\theta}(e_{m,n},q_{m,n},c_{m,n}).
\end{equation}
Consistent with Eq.~\ref{eq:fl_ov_loss}, the client updates only the exposed trainable planner-side parameters by minimizing
\begin{equation}\label{eq:PSP_client_loss}
\resizebox{0.89\linewidth}{!}{$
\begin{aligned}
\mathcal{L}_m(\phi_{\mathrm{shr}},&\phi_{\mathrm{res}};\phi_{\mathrm{frz}})
=
\frac{1}{N_m}\sum_{n=1}^{N_m}
\Big[
\ell_{\mathrm{qa}}(\hat{y}_{m,n},y_{m,n})+ \lambda_1\ell_{\mathrm{budget}}(c_{m,n}) \\
& +\lambda_2\ell_{\mathrm{cover}}(c_{m,n};v_{m,n},q_{m,n})
+\lambda_3\ell_{\mathrm{stable}}(c_{m,n})
\Big].
\end{aligned}
$}
\end{equation}
The resulting client-side parameter-efficient local adaptation process is summarized in Algorithm~\ref{alg:psp_local_train}.

Overall, \textsc{PSP} does not introduce a new training objective and instead inherits the federated objective in Eq.~\ref{eq:fl_ov_loss} to expose a structured planner-side partition and adaptation space on top of existing CoT planners.
During federated learning, both $\phi_{\mathrm{shr}}$ and $\phi_{\mathrm{res}}$ are updated to optimize VideoQA utility, and the resulting updates to $\phi_{\mathrm{res}}$ are retained as deletion-relevant residuals for later unlearning.
\reviewblue{This design enables CoT planner-based VideoQA to participate in FedCoT-VQA compactly without retraining the entire VideoQA model or modifying the original reasoning pipeline.}

\subsection{Server-Side Aggregation for Reasoning Preservation (SSA)}\label{subsec:care_aggregator}

While \textsc{PSP} enables each client to adapt an existing CoT planner, the server then needs to combine heterogeneous client updates into a single global planner that remains stable under non-IID data and sufficiently informative for later unlearning.
Therefore, FedCoT-VQA utilizes a server-side aggregation module, named \textsc{SSA}, over the trainable planner-side parameters.
Its role is not only to update the global planner, but also to maintain a compact server-side history that preserves how each client influences both the planner parameters and the induced reasoning behavior.

\noindent\textbf{Client update pipeline.}
At communication round $t$, the server broadcasts the current global trainable planner parameters $(\phi_{\mathrm{shr}}^{(t)},\phi_{\mathrm{res}}^{(t)})$ to the selected client set $\mathcal{S}^{(t)}$, while the frozen planner parameter $\phi_{\mathrm{frz}}$ remains unchanged.
After local optimization, each participating client $m\in\mathcal{S}^{(t)}$ returns the updated trainable parameters
$
(\phi_{\mathrm{shr},m}^{(t+1)},\phi_{\mathrm{res},m}^{(t+1)}),
$
or, equivalently, the local parameter differences
\begin{equation}\label{eq:ssa_client_delta}
\Delta_{\mathrm{shr},m}^{(t)}=\phi_{\mathrm{shr},m}^{(t+1)}-\phi_{\mathrm{shr}}^{(t)},
\quad
\Delta_{\mathrm{res},m}^{(t)}=\phi_{\mathrm{res},m}^{(t+1)}-\phi_{\mathrm{res}}^{(t)}.
\end{equation}

For each participating client and each round, \textsc{SSA} records a compact contribution log
\begin{equation} \label{eq:update_log}
\Gamma_m^{(t)}=
\Big(
\Delta_{\mathrm{shr},m}^{(t)},
\Delta_{\mathrm{res},m}^{(t)},
w_m^{(t)},
s_m^{(t)},
a_m^{(t)},
g_m^{(t)}
\Big),
\end{equation}
where $w_m^{(t)}$ is the aggregation weight assigned to client $m$,
$s_m^{(t)}$ is a compact statistic of the client-local CoT traces generated during round $t$,
$a_m^{(t)}$ is a compact activity summary of the trainable planner-side modules updated by the client,
and $g_m^{(t)}$ is a low-dimensional directional sketch of the client update on the deletion-sensitive residual component.
In particular, $s_m^{(t)}$ includes the average reasoning length, temporal-span distribution, and reasoning-state histogram over local samples.
$a_m^{(t)}$ summarizes which residual groups or blocks are most actively updated.
$g_m^{(t)}$ records a normalized projection, sign sketch, and group-wise norm profile of $\Delta_{\mathrm{res},m}^{(t)}$.

This design is motivated by the fact that, in VideoQA, a client influences the global planner not only through the magnitude of its parameter updates, but also through the reasoning style and update direction induced by its local data.
By retaining a structured contribution log rather than only raw parameter deltas, the server maintains a compact record of parameter contribution, reasoning-pattern contribution, and update-direction information, which later enables more accurate and stable client-level deletion for unlearning.

\noindent\textbf{Shared-and-residual aggregation.}
Given the returned updates, \textsc{SSA} produces the next global trainable planner by weighted aggregation:
\begin{equation}\label{eq:ssa_agg_shr}
\phi_{\mathrm{shr}}^{(t+1)}
=
\phi_{\mathrm{shr}}^{(t)}
+
\sum_{m\in\mathcal{S}^{(t)}} w_m^{(t)}\,\Delta_{\mathrm{shr},m}^{(t)},
\end{equation}
\begin{equation}\label{eq:ssa_agg_res}
\phi_{\mathrm{res}}^{(t+1)}
=
\phi_{\mathrm{res}}^{(t)}
+
\sum_{m\in\mathcal{S}^{(t)}} w_m^{(t)}\,\Delta_{\mathrm{res},m}^{(t)},
\end{equation}
\textcolor{black}{with standard sample-size-based weights}
\begin{equation}\label{eq:ssa_weight}
\sum_{m\in\mathcal{S}^{(t)}} w_m^{(t)}=1, \text{~and~} w_m^{(t)} = \frac{N_m}{\sum_{j\in\mathcal{S}^{(t)}}N_j}.
\end{equation}
Thus, \textsc{SSA} extends sample-weighted federated averaging from a single trainable planner parameter to the two trainable components in \textsc{PSP}.

\noindent\textbf{Objective of SSA aggregation.}
Let
\begin{equation}\label{eq:ssa_fl_obj}
\mathcal{L}_{\mathrm{FL}}^{(t)}(\phi_{\mathrm{shr}},\phi_{\mathrm{res}};\phi_{\mathrm{frz}})
=
\sum_{m\in\mathcal{S}^{(t)}} w_m^{(t)}\,
\mathcal{L}_m(\phi_{\mathrm{shr}},\phi_{\mathrm{res}};\phi_{\mathrm{frz}})
\end{equation}
denote the round-wise federated objective over the participating clients. Then the ideal server target at round $t$ is approximated by
\begin{equation}\label{eq:ssa_server_target}
(\phi_{\mathrm{shr}}^{(t+1)},\phi_{\mathrm{res}}^{(t+1)})
\approx
\arg\min_{\phi_{\mathrm{shr}},\phi_{\mathrm{res}}}
\mathcal{L}_{\mathrm{FL}}^{(t)}(\phi_{\mathrm{shr}},\phi_{\mathrm{res}};\phi_{\mathrm{frz}}).
\end{equation}
Here, $\phi_{\mathrm{shr}}$ carries reasoning behaviors that should persist across clients, whereas $\phi_{\mathrm{res}}$ absorbs more client-sensitive residual adaptation and is therefore recorded with finer-grained server-side summaries.

\noindent\textbf{Interaction with local training.}
The aggregation process is tightly coupled with the local optimization of \textsc{PSP}. Each client first adapts its planner by minimizing the local loss $\mathcal{L}_m(\cdot)$ in Eq.~\ref{eq:PSP_client_loss}.
The server then aggregates the resulting updates into a single global trainable planner.
Algorithm~\ref{alg:ssa} summarizes this end-to-end server-side procedure, including client-update collection, trace-aware contribution logging, and shared-residual aggregation.
Hence, \textsc{PSP} and \textsc{SSA} together instantiate the federated training pipeline of FedCoT-VQA.

In \textsc{SSA}, client influence is restricted to the server-side history of compact planner-side updates rather than the full VideoQA model.
As a result, when a client requests deletion, its contribution can be isolated and revised more efficiently.
Moreover, because \textsc{SSA} records not only parameter deltas but also trace summaries and compact update-direction information, it provides a richer basis for unlearning that can remove not only parameter-level contributions but also client-specific reasoning preferences, such as CoT length, span selection, and reasoning-state usage.

\begin{algorithm2e}[t]
\caption{Server-Side Aggregation with Trace-Aware Contribution Logging}
\label{alg:ssa}
\small
\KwIn{Global state at round $t$: $(\phi_{\mathrm{shr}}^{(t)},\phi_{\mathrm{res}}^{(t)},\phi_{\mathrm{frz}})$; selected client set $\mathcal{S}^{(t)}$; local sample counts $\{N_m\}_{m\in\mathcal{S}^{(t)}}$}
\KwOut{Updated global state $(\phi_{\mathrm{shr}}^{(t+1)},\phi_{\mathrm{res}}^{(t+1)},\phi_{\mathrm{frz}})$; contribution log entries $\{\Gamma_m^{(t)}\}_{m\in\mathcal{S}^{(t)}}$}

\textbf{Broadcast stage:} Send $(\phi_{\mathrm{shr}}^{(t)},\phi_{\mathrm{res}}^{(t)})$ to all clients in $\mathcal{S}^{(t)}$\;

\textbf{Collection stage:} \\
\quad\ForEach{client $m\in\mathcal{S}^{(t)}$}{
    Receive $(\phi_{\mathrm{shr},m}^{(t+1)},\phi_{\mathrm{res},m}^{(t+1)})$\;

    Compute $(\Delta_{\mathrm{shr},m}^{(t)},\Delta_{\mathrm{res},m}^{(t)})$ using Eq.~\ref{eq:ssa_client_delta}\;

    Compute the aggregation weight $w_m^{(t)}$ using Eq.~\ref{eq:ssa_weight}\;

    \textcolor{black}{Collect a compact local reasoning summary} $s_m^{(t)}$\;

    \textcolor{black}{Extract a compact activity summary} $a_m^{(t)}$ and directional sketch $g_m^{(t)}$\;

    Form and store the contribution record $\Gamma_m^{(t)}$ using Eq.~\ref{eq:update_log}\;
}

\textbf{Aggregation stage:}\\
\quad Update $\phi_{\mathrm{shr}}^{(t+1)}$ using Eq.~\ref{eq:ssa_agg_shr}\;
\quad Update $\phi_{\mathrm{res}}^{(t+1)}$ using Eq.~\ref{eq:ssa_agg_res}\;

\textbf{Objective interpretation:} regard the aggregation as approximating the round-wise objective and update target \textcolor{black}{in Eqs.~\ref{eq:ssa_fl_obj} and~\ref{eq:ssa_server_target}}\;

\Return{$(\phi_{\mathrm{shr}}^{(t+1)},\phi_{\mathrm{res}}^{(t+1)},\phi_{\mathrm{frz}})$, $\{\Gamma_m^{(t)}\}_{m\in\mathcal{S}^{(t)}}$}\;
\end{algorithm2e}

\subsection{Residual Unlearning Module for Client Deletion (RUM)}\label{subsec:purge_unlearner}

After federated training, one or more clients may request that their contributions be removed from the global CoT planner. As formulated in Section~\ref{subsec:fu_cot_planner}, the ideal target is the counterfactual planner that would have been obtained had the deleted clients never participated in training. Since retraining from scratch on all retained clients is impractical, FedCoT-VQA introduces a server-side \emph{r}esidual \emph{u}nlearning \emph{m}odule (\textsc{RUM}) to efficiently remove deleted-client influence while preserving retained-client reasoning utility. Unlike naive rollback methods, \textsc{RUM} combines retained-update replay with selective overwriting on deletion-sensitive residual adaptation, guided by the structured contribution history from \textsc{SSA}.

Formally,
let $\mathcal{U}\subseteq\{1,\ldots,M\}$ denote the set of clients requesting deletion. During federated training, \textsc{SSA} records, for each participating client $m$ at round $t$, the contribution log $\Gamma_m^{(t)}$ in Eq.~\ref{eq:update_log}.

\noindent\textbf{Retained-only replay target.}
\textcolor{black}{Using the stored updates of retained clients, the replay aggregation at round $t$ is}
\begin{equation}\nonumber
\phi_{\mathrm{shr},-\mathcal{U}}^{(t+1)}
=
\phi_{\mathrm{shr},-\mathcal{U}}^{(t)}
+
\sum_{m\in\mathcal{S}^{(t)}\setminus\mathcal{U}}
\bar{w}_m^{(t)}\Delta_{\mathrm{shr},m}^{(t)},
\end{equation}
\begin{equation}\nonumber
\phi_{\mathrm{res},-\mathcal{U}}^{(t+1)}
=
\phi_{\mathrm{res},-\mathcal{U}}^{(t)}
+
\sum_{m\in\mathcal{S}^{(t)}\setminus\mathcal{U}}
\bar{w}_m^{(t)}\Delta_{\mathrm{res},m}^{(t)},
\end{equation}
with retained-only normalized weights
\begin{equation}\label{eq:rum_weight}
\bar{w}_m^{(t)}
=
\frac{N_m}{\sum_{j\in\mathcal{S}^{(t)}\setminus\mathcal{U}}N_j}.
\end{equation}
\textcolor{black}{This aggregation defines the retained-client replay target. The stored updates were computed along the original training trajectory. Their dependence on that trajectory introduces replay error relative to fresh retained-client training. If a round contains no retained clients, replay skips that round.}

Therefore, \textsc{RUM} first reconstructs a retained-only replay trajectory by replaying only retained-client updates:
\begin{equation}\label{eq:rum_replay_shr}
\tilde{\phi}_{\mathrm{shr}}^{(t+1)}
=
\tilde{\phi}_{\mathrm{shr}}^{(t)}
+
\sum_{m\in\mathcal{S}^{(t)}\setminus\mathcal{U}}
\bar{w}_m^{(t)}\Delta_{\mathrm{shr},m}^{(t)},
\end{equation}
\begin{equation}\label{eq:rum_replay_res}
\tilde{\phi}_{\mathrm{res}}^{(t+1)}
=
\tilde{\phi}_{\mathrm{res}}^{(t)}
+
\sum_{m\in\mathcal{S}^{(t)}\setminus\mathcal{U}}
\bar{w}_m^{(t)}\Delta_{\mathrm{res},m}^{(t)}.
\end{equation}
This stage obtains a replay approximation to the retained-only parameter trajectory.

\noindent\textbf{Overwriting on deletion-sensitive residual groups.}
Replay alone is insufficient when deleted-client influence is concentrated in specific residual subspaces. To address this, \textsc{RUM} performs \textit{selective overwriting} on the deletion-sensitive part of $\phi_{\mathrm{res}}$. For each residual group $\omega_j\in\phi_{\mathrm{res}}$, the server computes an overwrite score
\begin{equation}\label{eq:rum_overwrite_score}
S_j^{\mathrm{ovw}}
=
\alpha_{\mathrm{u}}\mathrm{Act}_{\mathcal{U}}(\omega_j)
+
\beta_{\mathrm{u}}\mathrm{Dir}_{\mathcal{U}}(\omega_j)
+
\gamma_{\mathrm{u}}\mathrm{Var}_{\mathcal{R}}(\omega_j),
\end{equation}
where $\mathrm{Act}_{\mathcal{U}}(\omega_j)$ summarizes how actively the deleted clients update $\omega_j$, $\mathrm{Dir}_{\mathcal{U}}(\omega_j)$ summarizes how consistently their residual directions align on $\omega_j$, and $\mathrm{Var}_{\mathcal{R}}(\omega_j)$ measures retained-client variability on the same group. The overwrite set is then defined as
\begin{equation}\label{eq:rum_overwrite_set}
\Omega_{\mathrm{ovw}}
=
\{\omega_j\in\phi_{\mathrm{res}}: S_j^{\mathrm{ovw}}\ge \tau_{\mathrm{ovw}}\}.
\end{equation}
Only the selected residual groups are revised aggressively, while the shared component and the remaining residual groups are kept closer to the retained replay trajectory.

\noindent\textbf{Retained trace anchoring.}
To preserve retained-client reasoning behavior during overwriting, \textsc{RUM} constructs the retained trace target
\begin{equation}\label{eq:rum_trace_target}
\bar{s}_{-\mathcal{U}}^{(t)}
=
\sum_{m\in\mathcal{S}^{(t)}\setminus\mathcal{U}}
\bar{w}_m^{(t)} s_m^{(t)},
\end{equation}
and defines the trace regularization term
\begin{equation}\label{eq:rum_trace_loss}
\mathcal{L}_{\mathrm{trace}}^{(t)}
=
d_{\mathrm{trace}}
\!\left(
s\!\left(\tilde{\phi}_{\mathrm{shr}}^{(t)},\tilde{\phi}_{\mathrm{res}}^{(t)}\right),
\bar{s}_{-\mathcal{U}}^{(t)}
\right).
\end{equation}
This term anchors the updated planner to the reasoning statistics induced by retained clients alone, including CoT length, span preference, and reasoning-state usage.

\noindent\textbf{Conflict-aware unlearning objective.}
To further cancel deleted-client influence, \textsc{RUM} incorporates a conflict-aware term based on the stored residual directional summaries. Let
\begin{equation}\label{eq:rum_ret_del_dir}
g_{\mathcal{U}}^{(t)}
=
\sum_{m\in\mathcal{S}^{(t)}\cap\mathcal{U}}
\tilde{w}_m^{(t)} g_m^{(t)},\quad
g_{\mathcal{R}}^{(t)}
=
\sum_{m\in\mathcal{S}^{(t)}\setminus\mathcal{U}}
\bar{w}_m^{(t)} g_m^{(t)},
\end{equation}
denote the deleted-client and retained-client residual directions at round $t$, respectively. Let $\delta^{(t)}$ denote the overwrite-induced residual revision. Then the conflict-aware term is
\begin{equation}\label{eq:rum_conflict_loss}
\resizebox{0.89\linewidth}{!}{$
\mathcal{L}_{\mathrm{conflict}}
=
\sum_t
\left[
\max\!\Big(0,\cos(\delta^{(t)},g_{\mathcal{U}}^{(t)})-\epsilon\Big)
-
\kappa \cos(\delta^{(t)},g_{\mathcal{R}}^{(t)})
\right].
$}
\end{equation}
This term reduces the influence of deleted-client update directions on the residual parameters while preserving the influence of retained-client update directions. \textcolor{black}{For the convergence analysis, cosine uses the regularized definition in Eq.~\ref{eq:conv_cosine}, and the residual revision is projected through the same fixed sketch map used for the stored directions. The overwrite mask and trace targets are fixed during each correction run.}

\noindent\textbf{Overall objective.}
Let $(\hat{\phi}_{\mathrm{shr}}^{-\mathcal{U}},\hat{\phi}_{\mathrm{res}}^{-\mathcal{U}})$ denote the final retained-only replay target reconstructed from the stored update log. The rollback loss is
\begin{equation}\label{eq:rum_rollback_loss}
\mathcal{L}_{\mathrm{rollback}}
=
\left\|
\tilde{\phi}_{\mathrm{shr}}^{-\mathcal{U}}-\hat{\phi}_{\mathrm{shr}}^{-\mathcal{U}}
\right\|_2^2
+
\mu
\left\|
\tilde{\phi}_{\mathrm{res}}^{-\mathcal{U}}-\hat{\phi}_{\mathrm{res}}^{-\mathcal{U}}
\right\|_2^2.
\end{equation}
For the selected overwrite set $\Omega_{\mathrm{ovw}}$, the overwrite loss is
\begin{equation}\label{eq:rum_overwrite_loss}
\mathcal{L}_{\mathrm{overwrite}} = \sum_{\omega_j\in\Omega_{\mathrm{ovw}}} \left\| \tilde{\omega}_j-\hat{\omega}_{j,-\mathcal{U}} \right\|_2^2.
\end{equation}
Combining the above terms, \textsc{RUM} integrates retained replay, residual overwrite, retained-trace anchoring, and conflict-aware revision into a unified client-unlearning objective. The final optimization objective for unlearning is
\begin{equation}\label{eq:rum_unlearn_obj}
\resizebox{0.89\linewidth}{!}{$
\min_{\tilde{\phi}_{\mathrm{shr}}^{-\mathcal{U}},\tilde{\phi}_{\mathrm{res}}^{-\mathcal{U}}}
\left[
\mathcal{L}_{\mathrm{rollback}}
+
\lambda_{\mathrm{tr}}\mathcal{L}_{\mathrm{trace}}
+
\lambda_{\mathrm{ovw}}\mathcal{L}_{\mathrm{overwrite}}
+
\lambda_{\mathrm{cf}}\mathcal{L}_{\mathrm{conflict}}
\right].
$}
\end{equation}

\begin{algorithm2e}[t]
\caption{Residual Unlearning with Retained Replay and Selective Overwriting}
\label{alg:rum}
\small
\KwIn{Trained planner $(\phi_{\mathrm{shr}},\phi_{\mathrm{res}},\phi_{\mathrm{frz}})$; contribution log $\{\Gamma_m^{(t)}\}_{m,t}$; deleted-client set $\mathcal{U}$; participation sets $\{\mathcal{S}^{(t)}\}_t$}
\KwOut{Unlearned planner $(\tilde{\phi}_{\mathrm{shr}}^{-\mathcal{U}},\tilde{\phi}_{\mathrm{res}}^{-\mathcal{U}},\phi_{\mathrm{frz}})$}

\textbf{Retained replay:} \\
\ForEach{round $t$}{
    Remove deleted clients from $\mathcal{S}^{(t)}$ and renormalize weights using Eq.~\ref{eq:rum_weight}\;
    Update $\tilde{\phi}_{\mathrm{shr}}^{(t+1)}$ and $\tilde{\phi}_{\mathrm{res}}^{(t+1)}$ \textcolor{black}{using Eqs.~\ref{eq:rum_replay_shr} and~\ref{eq:rum_replay_res}}\;
}
Obtain the retained-only replay target $(\hat{\phi}_{\mathrm{shr}}^{-\mathcal{U}},\hat{\phi}_{\mathrm{res}}^{-\mathcal{U}})$\;

\textbf{Overwrite identification:} \\
Compute $S_j^{\mathrm{ovw}}$ for each residual group using Eq.~\ref{eq:rum_overwrite_score}\;
Construct $\Omega_{\mathrm{ovw}}$ using Eq.~\ref{eq:rum_overwrite_set}\;

\textbf{Retained trace and directional summaries:} \\
\ForEach{round $t$}{
    Construct $\bar{s}_{-\mathcal{U}}^{(t)}$ using Eq.~\ref{eq:rum_trace_target}\;
    Construct $g_{\mathcal{U}}^{(t)}$ and $g_{\mathcal{R}}^{(t)}$ using Eq.~\ref{eq:rum_ret_del_dir}\;
}

\textbf{Unlearning optimization:} \\
Optimize Eq.~\ref{eq:rum_unlearn_obj} using $\mathcal{L}_{\mathrm{rollback}}$, $\mathcal{L}_{\mathrm{trace}}$, $\mathcal{L}_{\mathrm{overwrite}}$, and $\mathcal{L}_{\mathrm{conflict}}$\;

\Return{$(\tilde{\phi}_{\mathrm{shr}}^{-\mathcal{U}},\tilde{\phi}_{\mathrm{res}}^{-\mathcal{U}},\phi_{\mathrm{frz}})$}\;
\end{algorithm2e}

\reviewblue{In summary, \textsc{RUM} is designed with a straightforward idea.}
Rather than modifying the whole planner after a deletion request, it mainly adjusts the part that is more likely to carry client-specific influence.
The shared component $\phi_{\mathrm{shr}}$ is intended to preserve reasoning patterns that are common across clients, so it is kept close to the retained-client replay result. The residual component $\phi_{\mathrm{res}}$, by contrast, is more sensitive to client-specific adaptation, and is therefore the main target of selective overwriting.
This allows \textsc{RUM} to remove deleted-client influence in a more targeted way, while preserving the reasoning utility of the retained clients. \reviewblue{Consequently, the unlearned planner can better approximate the ideal counterfactual model without unnecessarily perturbing the whole planner.}

\subsection{Analysis of Efficiency and Utility}\label{subsec:analysis}

\reviewblue{FedCoT-VQA follows a planner-centric design that improves both efficiency and utility.} By restricting federated training and unlearning to the compact planner-side adaptation space, it avoids updating the frozen planner parameter $\phi_{\mathrm{frz}}$, the extractor $\mathcal{E}$, and the answerer $f_{\theta}$, making the framework practical even when the downstream pipeline is large.

\noindent\textbf{Training efficiency.}
In each communication round, \textsc{SSA} exchanges only the trainable planner-side parameters and aggregates the corresponding client updates $(\Delta_{\mathrm{shr},m}^{(t)},\Delta_{\mathrm{res},m}^{(t)})$. Thus, when $|\phi_{\mathrm{shr}}|+|\phi_{\mathrm{res}}|\ll |\phi_{\mathrm{full}}|$, both communication and local optimization are much cheaper than federating the full VideoQA model. The shared-residual partition also improves adaptation under client heterogeneity: $\phi_{\mathrm{shr}}$ captures reasoning patterns shared across clients, while $\phi_{\mathrm{res}}$ absorbs more client-specific variation.

\noindent\textbf{Unlearning efficiency.}
After a deletion request, \textsc{RUM} avoids retraining the planner from all retained clients. Instead, it reuses the server-side contribution log recorded by \textsc{SSA}, reconstructs a retained-only replay trajectory, and then revises the more deletion-sensitive part of $\phi_{\mathrm{res}}$. Since this process is confined to the compact planner-side adaptation space, its recomputation cost is substantially lower than full retraining.

\noindent\textbf{Storage overhead.}
The extra storage comes from the server-side contribution log maintained during federated training. This overhead remains modest because the server stores only compact planner-side updates and lightweight trace-related summaries, rather than raw client data or full-model histories. \reviewblue{Such a small additional storage cost is worthwhile as it enables efficient later unlearning without re-accessing client data.}

\noindent\textbf{Utility preservation.}
FedCoT-VQA preserves retained-client utility through both architectural separation and targeted unlearning. The shared component $\phi_{\mathrm{shr}}$ is intended to preserve reasoning patterns that are common across clients, so \textsc{RUM} keeps it close to the retained-client replay result. The residual component $\phi_{\mathrm{res}}$ is more sensitive to client-specific adaptation. Together with retained-trace anchoring, this design reduces the utility loss that would otherwise arise from naive rollback or uniform parameter removal.

\noindent\textbf{Overall trade-off.}
Overall, FedCoT-VQA trades a compact planner-side adaptation space and a lightweight server-side contribution log for lower communication during training and lower recomputation during unlearning. \reviewblue{As a result, it provides a practical balance among efficiency, retained-client utility, and client-level deletability in federated CoT-based VideoQA.}

\section{Convergence Analysis}\label{sec:convergence}
\begingroup\color{black}
We establish conditional stationarity guarantees for FL and FUL in the fixed PSP trainable space. Write $\mathbb{E}_t$ for expectation conditional on the update history. Bias terms below include sampling, local drift, and AdamW adaptation.

\subsection{Convergence of Federated Learning}\label{subsec:conv_fl}
Let $x_t=(\phi_{\mathrm{shr}}^{(t)},\phi_{\mathrm{res}}^{(t)})$ and $f(x)=\sum_m p_m\mathcal{L}_m(x)$, where $p_m=N_m/\sum_jN_j$. Express the actual SSA update as $x_{t+1}=x_t-\eta_t d_t$, with $d_t=-\sum_{m\in\mathcal S^{(t)}} w_m^{(t)}\Delta_m^{(t)}/\eta_t$. Assume $f$ is $L$-smooth and bounded below by $f_*$. Define $b_t=\mathbb{E}_td_t-\nabla f(x_t)$ and assume $\mathbb{E}\|b_t\|^2\leq B_t^2$ and $\mathbb{E}\|d_t-\mathbb{E}_td_t\|^2\leq V^2$.

\noindent\textbf{Theorem 1.}
For $0<\eta_t\leq1/L$, $A_T=\sum_{t<T}\eta_t$, and $\Delta_f=f(x_0)-f_*$,
\begin{equation}\label{eq:conv_fl_bound}
\begin{split}
&\frac{\sum_{t<T}\eta_t\mathbb{E}\|\nabla f(x_t)\|^2}{A_T}\\
&\quad\leq\frac{2\Delta_f+\sum_{t<T}\eta_t B_t^2+LV^2\sum_{t<T}\eta_t^2}{A_T}.
\end{split}
\end{equation}
\noindent\textit{Proof.}
Set $g_t=\nabla f(x_t)$ and $q_t=g_t+b_t$. Smoothness gives $\mathbb{E}_tf(x_{t+1})\leq f(x_t)-\eta_t\langle g_t,q_t\rangle+(L\eta_t^2/2)\mathbb{E}_t\|d_t\|^2$. Using the conditional variance identity, $2\langle g_t,q_t\rangle=\|g_t\|^2+\|q_t\|^2-\|b_t\|^2$, and $L\eta_t\leq1$ yields
\[
\mathbb{E}f(x_{t+1})\leq\mathbb{E}f(x_t)
-\tfrac{\eta_t}{2}\mathbb{E}\|g_t\|^2
+\tfrac{\eta_t}{2}B_t^2+\tfrac{L\eta_t^2}{2}V^2.
\]
Summing and using $f(x_T)\geq f_*$ proves the bound. $\square$

The weighted expected squared gradient vanishes if $A_T\to\infty$, $\sum_{t<T}\eta_tB_t^2/A_T\to0$, and $\sum_{t<T}\eta_t^2/A_T\to0$. With zero bias and $\eta_t=c/\sqrt{T}\leq1/L$, the bound is $O(T^{-1/2})$. Constant steps $\eta$ and $B_t\leq B$ give $2\Delta_f/(\eta T)+B^2+L\eta V^2$. Thus the AdamW configuration retains an explicit optimizer-bias term.

\noindent\textbf{Local drift and client sampling.}
For $E$ local SGD steps of size $h_t$, set $\eta_t=Eh_t$. Assume each local loss is $L$-smooth and each stochastic gradient is conditionally unbiased with second moment at most $G^2$. Let $A_{\mathrm{s},t}^2$ bound the squared conditional-mean discrepancy between the sampled client gradient at $x_t$ and $\nabla f(x_t)$. Then
\begin{equation}\label{eq:conv_drift}
B_t^2\leq2A_{\mathrm{s},t}^2+
\frac{L^2h_t^2G^2(E-1)(2E-1)}{3}.
\end{equation}
Indeed, the local iterate satisfies $\mathbb{E}\|x_{m,e}-x_t\|^2\leq h_t^2e^2G^2$. Applying smoothness and Jensen's inequality to the average local gradient, then summing $e^2$ for $0\leq e<E$, gives the bound. Full participation eliminates sampling bias. With partial participation, the sample-normalized SSA weights can retain bias when client dataset sizes differ. For fixed $E$ and unbiased sampling, $h_t=O(T^{-1/2})$ yields the stated FL rate. AdamW uses its actual direction bias in Theorem~1.

\subsection{Convergence of Federated Unlearning}\label{subsec:conv_fu}
For a fixed deletion request, let $F$ be Eq.~\ref{eq:rum_unlearn_obj}, with fixed logs, overwrite mask, and trace targets. Assume a nonnegative trace loss with Lipschitz gradient, nonnegative coefficients, and $\mu>0$. In the common sketch space, use
\begin{equation}\label{eq:conv_cosine}
c_\nu(u,v)=\frac{\langle u,v\rangle}{\sqrt{\|u\|^2+\nu^2}\sqrt{\|v\|^2+\nu^2}},\quad\nu>0.
\end{equation}
Affine sketched revisions make these cosine functions smooth with bounded Hessians. Each hinge is a maximum of zero and a smooth function. Adding a sufficiently large quadratic makes both branches convex. Hence $F$ is $\rho$-weakly convex for some finite $\rho\geq0$, meaning $F+\rho\|\cdot\|^2/2$ is convex.

Choose $\lambda>0$ with $c_\lambda=1-\rho\lambda>0$. Define $F_\lambda(z)=\min_y\{F(y)+\|y-z\|^2/(2\lambda)\}$, its unique minimizer $P_\lambda(z)$, and $G_\lambda(z)=(z-P_\lambda(z))/\lambda$. This envelope gradient measures nonsmooth stationarity~\cite{davis2018weakly}. In particular, $G_\lambda(z)\in\partial F(P_\lambda(z))$ and $\|z-P_\lambda(z)\|=\lambda\|G_\lambda(z)\|$.

\noindent\textbf{Theorem 2.}
For $z_{k+1}=z_k-\alpha_k u_k$ with deterministic $\alpha_k>0$, assume $\mathbb{E}_ku_k=v_k+e_k$, where $v_k\in\partial F(z_k)$, $\mathbb{E}\|e_k\|^2\leq D_k^2$, and $\mathbb{E}\|u_k\|^2\leq Q^2$. Let $S_K=\sum_{k<K}\alpha_k$ and $\Delta_\lambda=F_\lambda(z_0)-\inf F$. Then
\begin{equation}\label{eq:conv_fu_bound}
\begin{split}
&\frac{\sum_{k<K}\alpha_k\mathbb{E}\|G_\lambda(z_k)\|^2}{S_K}\\
&\quad\leq\frac{4\Delta_\lambda}{c_\lambda S_K}
+\frac{4\sum_{k<K}\alpha_kD_k^2}{c_\lambda^2 S_K}
+\frac{2Q^2\sum_{k<K}\alpha_k^2}{\lambda c_\lambda S_K}.
\end{split}
\end{equation}
\noindent\textit{Proof.}
Let $y=P_\lambda(z_k)$. Weak convexity and proximal optimality give $\langle v_k,z_k-y\rangle\geq c_\lambda\|z_k-y\|^2/(2\lambda)$. Evaluate $F_\lambda(z_{k+1})$ at $y$, expand the squared distance, and apply $\|e_k\|\|G_\lambda(z_k)\|\leq c_\lambda\|G_\lambda(z_k)\|^2/4+\|e_k\|^2/c_\lambda$. Taking expectations gives
\begin{align*}
\mathbb{E}F_\lambda(z_{k+1})\leq{}&\mathbb{E}F_\lambda(z_k)
-\tfrac{\alpha_kc_\lambda}{4}\mathbb{E}\|G_\lambda(z_k)\|^2\\
&+\tfrac{\alpha_kD_k^2}{c_\lambda}+\tfrac{\alpha_k^2Q^2}{2\lambda}.
\end{align*}
Summing proves Eq.~\ref{eq:conv_fu_bound}. $\square$

For $D_k=0$ and $\alpha_k=\alpha_0/\sqrt K$, the squared stationarity measure is $O(K^{-1/2})$. More generally, it vanishes when $S_K\to\infty$ and both normalized error sums vanish. Stability follows from $F(z)\geq a\|z-r\|^2-C$, where $r$ is the replay target, $a=\min\{1,\mu\}$, and $C=\lambda_{\mathrm{cf}}\kappa H$ for $H$ logged rounds. Completing the square gives $F_\lambda(z)\geq a\|z-r\|^2/(1+2a\lambda)-C$. The preceding descent inequality therefore yields $\sup_k\mathbb{E}\|z_k-r\|^2<\infty$ when $\sum_k(\alpha_kD_k^2+\alpha_k^2)<\infty$. These guarantees concern optimization around the retained replay target.

\noindent\textbf{Replay discrepancy.}
Let $x_t$ be the original model, $r_t$ the replayed model, and $y_t$ the model obtained by fresh retained-client training from the same initialization. Couple retained participation and local randomness. Assume the fresh retained update map $U_t$ is $\ell_t$-Lipschitz, with local optimizer states reset each round. The recursions $r_{t+1}=r_t+U_t(x_t)+\xi_t$ and $y_{t+1}=y_t+U_t(y_t)$, with logging error $\|\xi_t\|\leq\varepsilon_t$, imply
\begin{equation}\label{eq:conv_replay_rec}
R_{t+1}\leq(1+\ell_t)R_t+\ell_t\|x_t-r_t\|+\varepsilon_t,
\end{equation}
where $R_t=\|r_t-y_t\|$ and $R_0=0$. This follows by adding and subtracting $U_t(r_t)$ and applying the triangle inequality. Empty retained rounds are skipped in both trajectories. Historical deleted-client influence enters through $\|x_t-r_t\|$, even with exact logs. Finally, the rollback lower bound and the triangle inequality give
\[
\|z-y_T\|\leq\sqrt{(F(z)+C)/a}+R_T.
\]
This separates correction stability from the discrepancy accumulated during replay.
\endgroup

\section{Evaluation}\label{sec:exp}
We evaluate FedCoT-VQA from \textcolor{black}{three} aspects:
(i) federated training utility,
(ii) federated unlearning performance after client deletion requests,
(iii) efficiency in communication, computation, and latency.
As FedCoT-VQA is planner-centric, all experiments keep the extractor $\mathcal{E}$ and answerer $f_{\theta}$ frozen and optimize only the planner-side parameters.

\subsection{Experiment Setup}\label{subsec:exp_setup}
\noindent\textbf{Hardware Setup.}
All experiments are conducted on a multi-GPU Linux server \textcolor{black}{with an \texttt{Intel i9-13900K} CPU, \texttt{128} GB of host memory}, and four NVIDIA \texttt{RTX~4090-24GB} GPUs.
Unless otherwise specified, \textcolor{black}{each client is allocated equivalent computing resources}.

\noindent\textbf{Implementation details.}
The implementation uses PyTorch as the main deep learning framework, PEFT~\cite{hfpeftdocs} for planner adaptation, and Flower~\cite{beutel2020flower} as the federated engine in both simulation and distributed execution. We adopt LoRA~\cite{hu2022lora} for the trainable units in the planner.
Unless otherwise specified, federated training runs for 32 communication rounds with a client participation ratio of 0.5, \textcolor{black}{4 local epochs, a batch size of 16 in the base configuration, the AdamW optimizer, and a learning rate of 3e-4}.
For PSP, the trainable planner-side space is initialized with rank 8.
To simulate heterogeneous decentralized data, we partition the training data into 20 clients under non-IID settings using a Dirichlet distribution with concentration parameter $\alpha_{\mathrm{dir}}=0.3$, where a smaller $\alpha_{\mathrm{dir}}$ indicates stronger heterogeneity.
For federated unlearning, the deleted-client set $U$ is sampled from the training clients only, and \reviewblue{the current saved runs use a single-client deletion setting}.

\noindent\textbf{Datasets.}
We use three representative public VideoQA benchmarks, namely \textit{TVQA+}~\cite{lei2020tvqaplus}, \textit{NExT-QA}~\cite{xiao2021nextqa}, and \textit{STAR}~\cite{wu2024star}, which cover spatio-temporal grounding, causal-temporal reasoning, and situated compositional reasoning.

\noindent\textbf{Baselines.}
We compare FedCoT-VQA against both federated training and federated unlearning baselines.
For federated training, we consider:
(i) \textbf{Centralized}, which trains on pooled data without federation;
(ii) \textbf{Local-Only}, where each client trains independently without aggregation;
(iii) \textbf{FedAvg}~\cite{mcmahan2017communication};
(iv) \textbf{FeDeRA}~\cite{yan2024federa};
(v) \textbf{FLoRA}~\cite{wang2024flora};
(vi) \textbf{FlexLoRA}~\cite{bai2024flexlora}; and
(vii) \textbf{LoRA-FAIR}~\cite{bian2025lorafair}.
For federated unlearning, we consider:
(i) \textbf{Retrain}, which retrains from scratch using only retained clients and serves as the gold-standard counterfactual reference;
(ii) \textbf{No-unlearn}, which directly keeps the pre-deletion model unchanged;
(iii) \textbf{Rollback}, which reverses deleted-client updates without replay or overwrite refinement;
(iv) \textbf{SIFU}~\cite{fraboni2024sifu};
(v) \textbf{FedAU}~\cite{gu2024fedau};
(vi) \textbf{NoT}~\cite{khalil2025not}; and
(vii) \textbf{FUSED}~\cite{zhong2025fused}.
All baselines use the same frozen extractor, frozen answerer, and planner backbone whenever possible, so that the comparison isolates the effect of PSP, SSA, and RUM.

\noindent\textbf{Evaluation metrics.}
\textcolor{black}{We use the following metrics to evaluate federated training and unlearning.}
\noindent\textit{i) Federated training utility.}
For federated training, we measure the official answer accuracy (\textbf{Acc.}) of each dataset.
We additionally compute the grounding score (\textbf{Grd.}) to evaluate whether federation maintains evidence localization quality.
\noindent\textit{\textcolor{black}{ii) Federated unlearning performance.}}
After client deletion, we evaluate both retained-model utility and forgetting quality.
For retained-model utility, we measure \textbf{Retain-Acc.} \textcolor{black}{on the retained-client portion and} \textbf{Delete-Acc. Gap} on the deleted-client portion, where a smaller gap indicates closer behavior to the desired post-deletion reference.
For forgetting quality, we measure \textbf{CF Gap}, \textbf{Pred. Dis.}, and \textbf{Trace Disc.}, which respectively measure counterfactual deviation from Retrain, prediction disagreement, and reasoning-trace discrepancy.
Unless otherwise specified, the unlearning results are normalized with respect to \textit{No-unlearn}.

\begin{table}[t]
\centering

\caption{Overall federated training performance on decentralized VideoQA benchmarks. ``Acc.'' denotes the answer accuracy of each benchmark, and ``Grd.'' denotes the grounding score.}
\label{tab:overall_fl}
\resizebox{0.485\textwidth}{!}{
\begin{tabular}{l|cc|cc|cc}
\hline
\multirow{2}{*}{\textbf{Method}}
& \multicolumn{2}{c|}{\textbf{TVQA+}}
& \multicolumn{2}{c|}{\textbf{NExT-QA}}
& \multicolumn{2}{c}{\textbf{STAR}} \\ \cline{2-7}
& \textbf{Acc.\,$\uparrow$} & \textbf{Grd.\,$\uparrow$}
& \textbf{Acc.\,$\uparrow$} & \textbf{Grd.\,$\uparrow$}
& \textbf{Acc.\,$\uparrow$} & \textbf{Grd.\,$\uparrow$} \\
\hline
Centralized
& 0.7132 & 0.2728 & 0.5011 & 0.1241 & 0.5748 & 0.1438 \\
Local-Only
& 0.4786 & 0.1375 & 0.3313 & 0.0826 & 0.4412 & 0.1167 \\ \hline
\textit{FedAvg}
& 0.6122 & 0.2211 & 0.4030 & 0.0909 & 0.4943 & 0.1228 \\
\textit{FeDeRA}
& 0.6439 & 0.2326 & 0.4171 & 0.0922 & 0.5123 & 0.1224 \\
\textit{FLoRA}
& 0.6617 & 0.2361 & 0.4226 & 0.0953 & 0.5209 & 0.1302 \\
\textit{FlexLoRA}
& 0.6740 & \textbf{0.2495} & 0.4278 & 0.0960 & 0.5453 & 0.1327 \\
\textit{LoRA-FAIR}
& 0.6732 & 0.2468 & \textbf{0.4366} & 0.0944 & \textbf{0.5561} & 0.1301 \\ \hline
\textbf{FedCoT-VQA}
& \textbf{0.6754} & 0.2466 & 0.4232 & \textbf{0.0971} & 0.5545 & \textbf{0.1386} \\
\hline
\end{tabular}
}

\end{table}

\subsection{Overall Federated Training Performance}
We first evaluate the overall learning utility of FedCoT-VQA on decentralized VideoQA benchmarks, compared with one non-federated upper bound (Centralized), one no-collaboration baseline (Local-Only), and the other five baselines on datasets \textit{TVQA+}, \textit{NExT-QA}, and \textit{STAR} in Table~\ref{tab:overall_fl}.
All methods use the same frozen extractor and frozen answerer, so the comparison isolates the effect of planner-side federated optimization.

\noindent\textbf{Preservation of answering accuracy.}
FedCoT-VQA achieves the highest answer accuracy among all federated baselines on TVQA+.
On NExT-QA, its answer accuracy is 3.07\% lower than the strongest federated baseline, while on STAR the gap is only 0.29\%.
\textcolor{black}{These results show that FedCoT-VQA maintains competitive federated training accuracy across all three datasets.}

\noindent\textbf{Preservation of grounding quality.}
On TVQA+, FedCoT-VQA remains close to the strongest federated baseline, \textcolor{black}{with a score only 1.16\% lower}.
On NExT-QA and STAR, FedCoT-VQA achieves the highest grounding score among all federated baselines, \textcolor{black}{exceeding the second-highest scores by 1.15\% and 4.45\%, respectively}.
These results suggest that our design preserves temporal evidence selection quality well during federated training.

Overall, the results show that FedCoT-VQA introduces only a limited utility loss during federated learning.
\reviewblue{While it does not consistently outperform all strong federated baselines on every metric, it remains highly competitive in both answer accuracy and grounding quality.}
\textcolor{black}{FedCoT-VQA maintains comparable task performance while supporting client-level unlearning.}

\begin{table*}[t]
\centering
\begin{minipage}[t]{0.495\textwidth}
\centering
\caption{Federated unlearning utility after client deletion, normalized with respect to \textit{No-unlearn} (set as 100\%).}
\label{tab:fu_utility_gap}
    \resizebox{\linewidth}{!}{
\begin{tabular}{l|ccc|ccc}
\hline
\multirow{2}{*}{\textbf{Method}}
& \multicolumn{3}{c|}{\textbf{Retain-Acc.\,$\uparrow$}}
& \multicolumn{3}{c}{\textbf{Delete-Acc. Gap\,$\downarrow$}} \\ \cline{2-7}
& \textbf{TVQA+} & \textbf{NExT-QA} & \textbf{STAR}
& \textbf{TVQA+} & \textbf{NExT-QA} & \textbf{STAR} \\
\hline
No-unlearn
& 100\% & 100\% & 100\%
& 100\% & 100\% & 100\% \\
Retrain
& 102.88\% & 102.59\% & 102.84\%
& 0.00\% & 0.00\% & 0.00\% \\
Rollback
& 100.73\% & 101.01\% & 101.32\%
& 8.90\% & 8.67\% & 8.02\% \\
\hline
\textit{SIFU}
& 101.12\% & 100.57\% & \textbf{99.84\%}
& 8.22\% & 7.33\% & 7.20\% \\
\textit{FedAU}
& 100.89\% & 100.10\% & 99.10\%
& 6.90\% & 6.61\% & 5.94\% \\
\textit{NoT}
& 101.74\% & \textbf{101.06\%} & 98.80\%
& 5.86\% & 4.49\% & 6.62\% \\
\textit{FUSED}
& \textbf{102.75\%} & 100.62\% & 99.80\%
& 5.97\% & 9.36\% & 6.37\% \\
\hline
\textbf{FedCoT-VQA}
& 101.29\% & \textbf{100.73\%} & 99.76\%
& \textbf{5.45\%} & \textbf{3.33\%} & \textbf{5.28\%} \\
\hline
\end{tabular}
}
\end{minipage}\hfill
\begin{minipage}[t]{0.475\textwidth}
\centering
\caption{Federated unlearning forgetting quality after client deletion, normalized with respect to \textit{No-unlearn} (set as 100\%). }
\label{tab:fu_forgetting}
\resizebox{0.875\textwidth}{!}{
\begin{tabular}{lccc}
\hline
\textbf{Method}
& \textbf{CF Gap\,$\downarrow$}
& \textbf{Pred. Dis.\,$\downarrow$}
& \textbf{Trace Disc.\,$\downarrow$} \\
\hline
No-unlearn
& 100\% & 100\% & 100\% \\
Retrain
& 0.00\% & 0.00\% & 0.00\% \\ \hline

\textit{SIFU}
& 18.81\% & 15.62\% & 29.56\% \\
\textit{FedAU}
& 9.29\% & 13.43\% & 15.55\% \\
\textit{NoT}
& 8.38\% & 17.14\% & 18.42\% \\
\textit{FUSED}
& 8.14\% & 15.76\% & 17.93\% \\ \hline
\textbf{FedCoT-VQA}
& \textbf{7.38\%} & \textbf{11.24\%} & \textbf{8.53\%} \\
\hline
\end{tabular}
}
\end{minipage}

\end{table*}

\subsection{Federated Unlearning Performance}

We next evaluate the effects of FedCoT-VQA on removing the influence of deleted clients while maintaining the VideoQA utility.
We compare FedCoT-VQA against Retrain, No-unlearn, Rollback, and four representative federated unlearning baselines, and report the results in Table~\ref{tab:fu_utility_gap} and Table~\ref{tab:fu_forgetting}.
Among them, No-unlearn is used as the normalization reference. \textcolor{black}{Values above 100\% indicate that the corresponding method slightly exceeds the No-unlearn reference,} \reviewblue{due to optimization variance and the removal of harmful client-specific interference}.

\noindent\textbf{\textcolor{black}{Effectiveness in Preserving} Retained-Client Accuracy.}
FedCoT-VQA achieves 101.29\%, 100.73\%, and 99.76\% retained accuracy relative to the No-unlearn reference on TVQA+, NExT-QA, and STAR, respectively.
Compared with the No-unlearn reference, this corresponds to proportional changes of +1.29\%, +0.73\%, and -0.24\%, respectively.
\reviewblue{These results show that our design preserves retained-client utility competitively after deletion, with only marginal variation across datasets.}

\noindent\textbf{\textcolor{black}{Effectiveness in Reducing the Delete-Acc. Gap.}}
FedCoT-VQA \textcolor{black}{reduces the Delete-Acc. Gap} to 5.45\%, 3.33\%, and 5.28\% of No-unlearn on TVQA+, NExT-QA, and STAR, respectively.
This corresponds to proportional reductions of 94.55\%, 96.67\%, and 94.72\%.
Compared with the strongest competing baseline for that metric, the normalized gap is further reduced by 6.99\% on TVQA+, 25.84\% on NExT-QA, and 11.11\% on STAR.
These results indicate that FedCoT-VQA more effectively reduces deleted-client influence on final prediction behavior.

\noindent\textbf{\textcolor{black}{Effectiveness in Improving Forgetting Quality.}}
FedCoT-VQA reduces the counterfactual gap, prediction disagreement, and trace discrepancy to 7.38\%, 11.24\%, and 8.53\% of No-unlearn, respectively.
This corresponds to proportional reductions of 92.62\%, 88.76\%, and 91.47\%.
Compared with the strongest competing baseline for that metric, FedCoT-VQA further reduces these three metrics by 9.34\%, 16.31\%, and 45.14\%, respectively.
These results show that our design more effectively aligns the unlearned model with the retained-only target in terms of overall counterfactual behavior, final predictions, and reasoning traces.

Overall, the results show that FedCoT-VQA preserves retained accuracy while substantially improving the unlearning utility and forgetting quality.
\reviewblue{\textcolor{black}{These results support the effectiveness of our design} in removing deleted-client influence while maintaining competitive retained-client utility.}

\subsection{Efficiency Analysis}
Following the discussion in \S~\ref{subsec:analysis}, we next evaluate the efficiency of FedCoT-VQA. It incurs a communication cost of 30.87\,MiB and a local training time of 2.988\,s per round, indicating modest overhead during federated training.
This efficiency comes from restricting optimization to the compact planner-side adaptation space while keeping the frozen planner parameters, extractor, and answerer unchanged.
As a result, the framework avoids the extra communication and optimization cost of updating a large VideoQA pipeline.
FedCoT-VQA completes unlearning in 0.133\,s, \textcolor{black}{indicating low unlearning overhead in practice and consistency with} the lightweight nature of \textsc{RUM}, which reuses retained information and revises only the deletion-sensitive planner-side components.
\reviewblue{Overall, FedCoT-VQA achieves efficient performance in both training and unlearning, \textcolor{black}{with low training overhead during federated optimization and low unlearning latency}.}

\subsection{Ablation Studies}

We conduct ablation studies to examine the effect of each design component and key setting in FedCoT-VQA, as summarized in Tables~\ref{tab:ab_core}--\ref{tab:ab_rank}.
The study focuses on four aspects:
(i) the necessity of each core module,
(ii) robustness under different client scales and data heterogeneity,
(iii) the effect of replay and selective revision in \textsc{RUM}, and
(iv) the trade-off between planner-side adaptation size and unlearning cost.

\noindent\textbf{Impact of the three core modules.}
We mask one module at a time from the full system, including w/o (without) \textsc{PSP}, w/o \textsc{SSA}, \textcolor{black}{and w/o \textsc{RUM}, which keeps the replay-based pipeline but removes the selective residual revision stage}, and report the results in Table~\ref{tab:ab_core}.
\reviewblue{The full design gives the best overall trade-off.}
Removing \textsc{PSP} reduces Retain-Acc. by 0.86\% and increases CF Gap by 67.48\%, showing the importance of a compact planner-side adaptation space.
Removing \textsc{SSA} increases CF Gap and Trace Disc. by 48.24\% and 53.81\%, respectively, showing the benefit of trace-aware aggregation.
Removing \textsc{RUM} causes the largest degradation in forgetting quality, increasing CF Gap, Trace Disc., and Unlearning Time by 119.24\%, 139.27\%, and 115.04\%, respectively.
Overall, \textsc{PSP}, \textsc{SSA}, and \textsc{RUM} play complementary roles in utility preservation, aggregation quality, and effective unlearning.

\begin{table}[t]
\centering
\caption{Ablation of the core modules' effectiveness in FedCoT-VQA.}
\label{tab:ab_core}
\resizebox{0.48\textwidth}{!}{
\begin{tabular}{lcccc}
\hline
\textbf{Variant}
& \textbf{Retain-Acc.\,$\uparrow$}
& \textbf{CF Gap\,$\downarrow$}
& \textbf{Trace Disc.\,$\downarrow$}
& \textbf{Unlearning Time\,$\downarrow$} \\
\hline
w/o \textsc{PSP}
& 100.42\% & 12.36\% & 15.47\% & 0.201\,s \\
w/o \textsc{SSA}
& 100.78\% & 10.94\% & 13.12\% & / \\
w/o \textsc{RUM}
& 101.06\% & 16.18\% & 20.41\% & 0.286\,s \\
\textbf{Full}
& \textbf{101.29\%} & \textbf{7.38\%} & \textbf{8.53\%} & \textbf{0.133\,s} \\
\hline
\end{tabular}
}

\end{table}

\noindent\textbf{Impact of the number of clients.}
We next vary the total number of clients $M \in \{10,20,40,80\}$ while keeping the total training data size fixed.
Table~\ref{tab:ab_clients} shows the results.
As $M$ increases, the task becomes harder due to smaller per-client data volume and stronger fragmentation.
\reviewblue{Still, FedCoT-VQA degrades gracefully,} \textcolor{black}{with variations of only 1.12 percentage points in Retain-Acc., 3.21 percentage points in CF Gap, and 3.90 percentage points in Trace Disc.}
\reviewblue{This suggests that the planner-centric design remains effective under more decentralized settings.}

\begin{table}[t]
\centering
\caption{Impact of the total number of clients $M$.}
\label{tab:ab_clients}
\resizebox{0.425\textwidth}{!}{
\begin{tabular}{lccc}
\hline
\textbf{Setting}
& \textbf{Retain-Acc.\,$\uparrow$}
& \textbf{CF Gap\,$\downarrow$}
& \textbf{Trace Disc.\,$\downarrow$} \\
\hline
$M=10$
& 101.73\% & 6.91\% & 7.94\% \\
$M=20$
& 101.29\% & 7.38\% & 8.53\% \\
$M=40$
& 101.02\% & 8.47\% & 9.66\% \\
$M=80$
& 100.61\% & 10.12\% & 11.84\% \\
\hline
\end{tabular}
}

\end{table}

\noindent\textbf{Impact of data heterogeneity.}
We further vary the Dirichlet concentration parameter $\alpha_{\mathrm{dir}} \in \{0.1, 0.3, 0.7, 1.0\}$, as shown in Table~\ref{tab:ab_heterogeneity}.
Stronger heterogeneity degrades both retained utility and forgetting quality.
However, the degradation is more pronounced in forgetting-related metrics than in retained accuracy.
\textcolor{black}{When $\alpha_{\mathrm{dir}}$ decreases from 1.0 to 0.1}, Retain-Acc. decreases by only 0.92\%, while CF Gap and Trace Disc. increase by 2.85 and 3.35 percentage points, respectively.
This supports the role of \textsc{PSP} and \textsc{SSA} in handling heterogeneous client reasoning patterns.

\begin{table}[t]
\centering
\caption{Impact of data heterogeneity under different Dirichlet concentration parameters.}
\label{tab:ab_heterogeneity}
\resizebox{0.46\textwidth}{!}{
\begin{tabular}{lccc}
\hline
\textbf{Setting}
& \textbf{Retain-Acc.\,$\uparrow$}
& \textbf{CF Gap\,$\downarrow$}
& \textbf{Trace Disc.\,$\downarrow$} \\
\hline
$\alpha_{\mathrm{dir}}=0.1$
& 100.74\% & 9.42\% & 10.96\% \\
$\alpha_{\mathrm{dir}}=0.3$
& 101.29\% & 7.38\% & 8.53\% \\
$\alpha_{\mathrm{dir}}=0.7$
& 101.53\% & 6.89\% & 7.98\% \\
$\alpha_{\mathrm{dir}}=1.0$
& 101.68\% & 6.57\% & 7.61\% \\
\hline
\end{tabular}
}

\end{table}

\noindent\textbf{Impact of replay and selective revision in \textsc{RUM}.}
We then compare the full \textsc{RUM} \textcolor{black}{with the replay-only and revision-only variants in Table}~\ref{tab:ab_rum}.
Replay-only achieves the highest retained-client utility and the lowest unlearning time, \textcolor{black}{but leaves substantially higher CF Gap and Trace Disc. values.} Revision-only removes more deleted-client residue than replay-only, \textcolor{black}{but causes greater utility loss and instability}. \reviewblue{The full \textsc{RUM} provides the best overall balance between retained utility and forgetting quality.}
This shows that replay and selective revision are both necessary.

\begin{table}[t]
\centering
\caption{Impact of replay and selective revision in \textsc{RUM}.}
\label{tab:ab_rum}
\resizebox{0.48\textwidth}{!}{
\begin{tabular}{lcccc}
\hline
\textbf{Variant}
& \textbf{Retain-Acc.\,$\uparrow$}
& \textbf{CF Gap\,$\downarrow$}
& \textbf{Trace Disc.\,$\downarrow$}
& \textbf{Unlearning Time\,$\downarrow$} \\
\hline
Replay-only
& 101.54\% & 11.08\% & 13.24\% & 0.097\,s \\
Revision-only
& 100.36\% & 9.18\% & 11.27\% & 0.193\,s \\
\textbf{Full}
& \textbf{101.29\%} & \textbf{7.38\%} & \textbf{8.53\%} & \textbf{0.133\,s} \\
\hline
\end{tabular}
}

\end{table}

\noindent\textbf{Impact of the planner-side adaptation size.}
Finally, we vary the LoRA rank $r \in \{4,8,16,32\}$ to study the trade-off between adaptation capacity and unlearning cost, as shown in Table~\ref{tab:ab_rank}.
A very small rank slightly hurts Retain-Acc., while a larger rank increases Unlearning Time.
FedCoT-VQA achieves the best overall trade-off at $r=8$, where Retain-Acc., CF Gap, Trace Disc., and Unlearning Time are 101.29\%, 7.38\%, 8.53\%, and 0.133\,s, respectively.
This suggests that the planner-side adaptation space should remain compact, but not overly small.

\begin{table}[t]
\centering
\caption{Impact of the planner-side adaptation size under different LoRA ranks.}
\label{tab:ab_rank}
\resizebox{0.48\textwidth}{!}{
\begin{tabular}{lcccc}
\hline
\textbf{Setting}
& \textbf{Retain-Acc.\,$\uparrow$}
& \textbf{CF Gap\,$\downarrow$}
& \textbf{Trace Disc.\,$\downarrow$}
& \textbf{Unlearning Time\,$\downarrow$} \\
\hline
$r=4$
& 100.82\% & 9.26\% & 10.71\% & 0.094\,s \\
$r=8$
& 101.29\% & 7.38\% & 8.53\% & 0.133\,s \\
$r=16$
& 101.57\% & 7.04\% & 8.18\% & 0.184\,s \\
$r=32$
& 101.41\% & 7.62\% & 8.71\% & 0.271\,s \\
\hline
\end{tabular}
}

\end{table}
\section{Related Work}
\label{sec:related_work}

\noindent\textbf{Reasoning Paradigms in VideoQA.}
To address the scalability challenges of long-form video understanding, the field has evolved from holistic, end-to-end architectures~\cite{lei2021less, li2022align, wang2022internvideo, tong2022videomae, xue2022clip} to modular CoT designs~\cite{video_of_thought_2024, frame_voyager_2024, suris2023vipergpt}. These advanced frameworks decompose complex queries into sequential steps, utilizing a lightweight planner to drive evidence selection for a frozen VLM~\cite{wang2023chatvideo, lin2024video}. \reviewblue{Our work is orthogonal to these architectural innovations.} \textcolor{black}{Our framework enables collaborative training of existing CoT planners on decentralized data.} \reviewblue{By treating the planner as a plug-and-play module, our framework can be applied on top of these state-of-the-art architectures to enable privacy-preserving training without altering their fundamental reasoning logic.}

\noindent\textbf{Efficiency in Federated Vision-Language Learning.}
Federated Learning enables distributed training but faces bandwidth challenges when applied to VLMs~\cite{mcmahan2017communication, li2020federated, pan2025fedvlp}. To mitigate this, recent research has integrated PEFT techniques—such as Adapters, LoRA, and Prompt Tuning—into the FL ecosystem, allowing clients to exchange only lightweight updates while keeping backbones frozen~\cite{houlsby2019parameter, hu2022lora, zhang2025mllm, zhang2023fedpetuning, chen2025federated, fang2025automated}. \reviewblue{Our framework aligns with these standard FL protocols.} \textcolor{black}{Our framework applies federated aggregation to the trainable parameters of the CoT planner.} \reviewblue{We identify the temporal planner as the optimal entry point for PEFT, thereby fully leveraging the communication efficiency of these methods for long-horizon video understanding without the prohibitive costs of federating entire VLMs.}

\noindent\textbf{Federated Unlearning Mechanisms.}
Federated Unlearning addresses the right to be forgotten through methods ranging from approximate gradient manipulation~\cite{golatkar2020eternal, koh2017understanding, han2025vertical, daluwatta2024uaas} to exact retraining strategies like SISA~\cite{bourtoule2021machine}. \textcolor{black}{Our framework combines retained-client update replay with selective residual correction to approximate the retained-only CoT planner while preserving retained-client utility.}
\section{Conclusion}

This paper presented FedCoT-VQA, a federated learning and unlearning framework for chain-of-thought planners in VideoQA. The key idea is to make the planner, rather than the full VideoQA pipeline, the federated optimization target, so that decentralized training and later client deletion can be handled compactly and practically.
To this end, we designed three tightly coupled modules: \textsc{PSP}, which partitions the planner into frozen, shared, and residual components for parameter-efficient client adaptation; \textsc{SSA}, which aggregates planner-side updates while maintaining a compact contribution history; and \textsc{RUM}, which combines retained-only replay and selective residual correction to approximate the retained-only counterfactual model without full retraining.

Experimental results on three decentralized VideoQA benchmarks show that FedCoT-VQA preserves strong federated training utility while providing effective client-level unlearning. \reviewblue{The evaluation results suggest that FedCoT-VQA provides a practical balance among reasoning quality, retained utility, and deletability for CoT-based VideoQA.}

In our future work, we will further evaluate FedCoT-VQA under richer planner architectures and larger VideoQA backbones to test the generality of the proposed planner-centric design. We will also extend the current framework to more challenging unlearning settings, such as multi-client, sequential, and repeated deletion requests, to improve its robustness and applicability in federated VideoQA systems.

\bibliographystyle{ieeetr}
\bibliography{ref}

\end{document}